\documentclass[
reprint,
superscriptaddress,
amsmath,amssymb,
aps,prl,
]{revtex4-2}

\usepackage{graphicx}
\usepackage{dcolumn}
\usepackage{bm}
\usepackage[hidelinks]{hyperref}
\usepackage{xcolor}
\usepackage{soul}
\usepackage{physics}
\usepackage[acronym]{glossaries}
\usepackage{capt-of}
\usepackage[capitalise]{cleveref}
\glsdisablehyper

\newacronym{md}{MD}{Molecular Dynamics}
\newacronym{ikp}{IKP}{Improved Kelbg Potential}
\newacronym{pic}{PIC}{Particle-In-Cell}
\newacronym{bh}{BH}{Bethe-Heitler}
\newacronym{h}{H}{Haug}
\newacronym{s}{S}{Schwinger}
\newacronym{mj}{MJ}{Maxwell-J\"uttner}
\newacronym{lpm}{LPM}{Landau-Pomeranchuk-Migdal}
\newacronym{lw}{LW}{Li{\'e}nard–Wiechert}
\newacronym{ee}{$e$-$e$}{electron-electron}
\newacronym{ei}{$e$-$i$}{electron-ion}
\newacronym{ii}{$i$-$i$}{ion-ion}
\glsunset{ii}

\crefname{appendix}{Appendix}{Appendices}
\Crefname{appendix}{Appendix}{Appendices}

\begin{document}

\title{\textbf{Radiation in relativistic plasmas via particle simulations}}%

\author{Efstratios M. Kritikos}
\email{efstratios.kritikos@physics.ox.ac.uk}

\affiliation{Department of Physics, University of Oxford, Parks Road, Oxford OX1 3PU, United Kingdom}

\author{Pontus Svensson}

\affiliation{Institute of Radiation Physics, Helmholtz-Zentrum Dresden-Rossendorf (HZDR), D-01328 Dresden, Germany}

\author{Daniel Plummer}

\affiliation{Department of Physics, University of Oxford, Parks Road, Oxford OX1 3PU, United Kingdom}

\author{Brian Reville}

\affiliation{Max-Planck-Institut für Kernphysik (MPIK), Saupfercheckweg, Heidelberg D-69117, Germany}

\author{Bob Bingham}

\affiliation{Central Laser Facility, STFC Rutherford Appleton Laboratory, Didcot, OX11 0QX, United Kingdom}

\author{Gianluca Gregori}

\affiliation{Department of Physics, University of Oxford, Parks Road, Oxford OX1 3PU, United Kingdom}

\date{August 25, 2026}

\begin{abstract}
We investigate radiation in relativistic electron-proton isotropic plasmas using molecular dynamics, with velocity-dependent inertia, retarded interactions, and quantum-corrected pair potentials. Radiation spectra from the Li{\'e}nard–Wiechert fields agree with analytical models. Successive scattering events suppress bremsstrahlung even at weak coupling, while jitter radiation becomes comparable at higher coupling for all resolved frequencies. This first-principles framework will help develop radiation models for suppression and field effects that are inaccessible in typical particle-in-cell methods.
\end{abstract}

\maketitle

Relativistic plasmas are found in astrophysical environments, such as neutron-star atmospheres, gamma-ray bursts, and black-hole accretion disks, as well as laboratory facilities, such as inertial confinement fusion experiments, particle accelerators, and high-intensity lasers \cite{Tallents2022}. Radiation in these plasmas arises from particle-particle (bremsstrahlung) and particle-wave interactions \cite{Rybicki1985}. Bremsstrahlung originates from the acceleration of a charged particle, when deflected by another charged particle, leading to photon emission. At relativistic conditions, both \gls{ei} and \gls{ee} bremsstrahlung become comparable \cite{Haug1989}. In high-density plasmas, successive scattering leads to a suppression in the emitted bremsstrahlung spectrum, known as the \gls{lpm} effect \cite{Baier1998}. Radiation is also emitted from the interaction of particles with plasma-induced electromagnetic fields, namely synchrotron \cite{Schwinger1949} and its small deflection-angle limit \cite{Landau1975,ToptyghinFleishman} also called jitter radiation \cite{Medvedev2000}. The synchrotron limit occurs when the particles' deflection angle in the background electromagnetic field exceeds the beaming angle ($1/\gamma$) of the emitted radiation, where $\gamma$ is the Lorentz factor of the particle. Jitter occurs when the deflection angle is less than the beaming angle, thus requiring substantial power in sub-Larmor scale electromagnetic fluctuations. 

\gls{pic} simulations of relativistic plasmas (e.g., \cite{Vyskocil2018}) typically model bremsstrahlung using the theoretical models of \gls{bh} \cite{Bethe1934} and \gls{h} \cite{Haug1989} (or their tabulated values in Ref. \cite{Seltzer1985}). These models apply perturbative quantum electrodynamics and the first approximation of Born's collision theory to derive the differential photon emission cross-section arising respectively from \gls{ei} and \gls{ee} collisions. An important caveat is that these models assume isolated particle interactions and thus are not able to capture the \gls{lpm} effect \cite{Migdal1956}, which can become important in high-density plasma environments. Therefore, the validity of neglecting the \gls{lpm} phenomenon in relativistic plasmas needs to be assessed for different coupling constants $\Gamma = e^2/(4\pi \epsilon_0 a \kappa T)$ and relativistic parameters $\varepsilon = \kappa T/(m_e c^2)$. Here, $e$ and $m_e$ are the electron charge and electron mass, $\epsilon_0$ is the vacuum permittivity, $\kappa$ is the Boltzmann constant, $T$ is the plasma temperature, and $c$ is the speed of light. Moreover, $a=[3/(4 \pi n)]^{1/3}$, is the Wigner-Seitz radius, where $n$ is the total number density.

In \gls{pic} simulations, particles represent a collection of physical particles, modelled with spline shape functions \cite{Arber2015}. Consequently, the particles cannot generate or interact with electromagnetic fields at wavelengths shorter than the particle size. Thus, jitter radiation from such high-$\bm{k}$ fluctuations is essentially neglected. Studies on jitter radiation using \gls{pic} simulations (e.g., Refs.~\cite{Hededal,Sironi2009}) compute the \gls{lw} fields in a post-processing step after the simulation. This step is necessary since the incoherent synchrotron and jitter emission happens at wavelengths that are typically much smaller than the grid spacing \cite{Rybicki1985,Medvedev2000}. The post-processing, however, assumes a collisionless plasma so as not to include acceleration arising from bremsstrahlung and neglects the impact of particles interacting with and generating high $\bm{k}$-mode magnetic fields. A numerical framework is thus needed to compute microscale jitter radiation in collisional relativistic plasmas from first principles.

In this study, we investigate radiation in relativistic plasmas using a \gls{md} framework. By modelling individual particle-particle and particle-wave interactions, we retain a high degree of accuracy of the microscale interactions. Electrostatic \gls{md} has recently been used \cite{Kinney2025a} to measure bremsstrahlung in nonrelativistic plasmas with the simplified \gls{lw} fields, showing good agreement with the theory. To our knowledge, an \gls{md} approach to compute radiation in relativistic plasmas has not been developed yet.

The objectives of this work are to (i) establish a new numerical framework based on \gls{md} to compute radiation in relativistic plasmas; (ii) investigate bremsstrahlung and the impact of the \gls{lpm} phenomenon; and (iii) study radiation arising from particle-electromagnetic field interactions. To address these objectives, we start with the relativistic \gls{md} code recently developed in Ref. \cite{Svensson2024}, which is based on the LAMMPS software \cite{Thompson2022}. This framework uses relativistic kinematics, where electromagnetic interactions are separated into short-range electrostatic and long-range causally retarded interactions. For the long-range interactions, particles are modelled as Gaussian functions with width $\sigma$, which is set to $1.5a$. The resulting electromagnetic fields are then evolved on a grid, based on a generalization of the PPPM method \cite{Hockney2021}. It is noted that the induced, fluctuating electromagnetic fields remain within the domain, as they cannot escape from the periodic boundaries. Removing outgoing waves requires absorbing boundary conditions \cite{Kritikos2026}, which have considerable complexity and thus will be implemented in future work. The interaction originating from the difference between the Gaussian and point particle charge distribution is short-range but rapidly fluctuating. This short-range contribution is treated electrostatically (cf. Section S2 for more details). The code is extended to account for short-range quantum diffraction through the \gls{ikp} \cite{Filinov2004} and to compute radiation from the \gls{lw} fields \cite{Rybicki1985}. Thus, both electrons and ions are modelled as classical particles, while short-range quantum effects are captured via the \gls{ikp}. Moreover, we mainly focus on simulating plasmas within the classical regime where quantum degeneracy effects are negligible.

For each \gls{ee} and \gls{ei} pair the \gls{ikp} force is given by
\begin{equation}
\bm{F}_{I,ij}\!= \!\frac{k_c q_i q_j \bm{r}_{i j}}{r_{i j}^3} \left[ 1\!-\!\frac{\lambda_{ij}^2+2 r_{i j}^2}{\lambda_{ij}^2} e^{- \left(\frac{r_{i j}}{\lambda_{ij}} \right)^2}\!+\!\frac{2 r_{i j}^2}{\lambda_{ij}^2} e^{- \left(\frac{\chi r_{i j}}{\lambda_{ij}} \right)^2} \right],
\label{eq:ikp_f}
\end{equation}
where $k_c=1/(4 \pi \epsilon_0)$ is the Coulomb constant, $q_i$ is the charge of the $i^{\rm th}$ particle, $r_{ij}$ is the interparticle distance, $\lambda_{ij}^{2} = \hbar^2/(2\kappa T \mu_{ij})$ is the thermal de Broglie wavelength, $\mu_{ij}$ is the reduced mass, and the $\chi$ correction parameter for \gls{ee} and \gls{ei} interactions is given in \cref{app:ikp}. \gls{ikp} recovers the Coulomb potential at large separations. The \gls{ii} interactions are treated as Coulombic, i.e., setting $\lambda_{ij} \rightarrow 0$ in \cref{eq:ikp_f}. The \gls{ikp} was derived from the nonrelativistic two-particle Bloch equation \cite{Filinov2004}, but it has also been applied to relativistic quark-gluon plasmas \cite{Filinov2015}, as no relativistic generalisation exists yet. A possible correction would be to use the relativistic $\lambda_{ij}$ to account for relativistic changes in the wavelength \cite{Hao2023}. However, the relativistic and nonrelativistic $\lambda_{ij}$ are similar near our simulation conditions at $\varepsilon \le 1$ (cf. Fig.~S2), and thus the non-relativistic value is used.

Radiation in relativistic plasmas is here computed from the \gls{lw} fields. Specifically, the total energy per unit angular frequency $\omega$ and per unit solid angle $\Omega$ that is radiated from a collection of particles at the emission time $t$ is computed from the Fourier transform of the \gls{lw} electric field as \cite{Rybicki1985}
\begin{equation}
\frac{\dd^2 I}{\dd \Omega \dd \omega} = \left|\int_{- \infty}^{\infty} \sum_j \frac{q_j\bm{n} \times \{(\bm{n} - \bm{\beta}_j) \times \dot{\bm{\beta}}_j \}}{4 (\pi^3 \epsilon_0 c)^{1/2} (1 - \bm{n} \cdot \bm{\beta}_j)^2} e^{i \omega \left( t - \bm{n} \cdot \frac{\bm{r}_j}{c} \right)} \dd t \right|^2,
\label{eq:lw}
\end{equation}
where the acceleration-force relation $\dot{\bm{\beta}}_j = [\bm{F}_j - (\bm{F}_j \cdot \bm{\beta}_j) \bm{\beta}_j]/(m c \gamma)$ is used here since particle force $\bm{F}$ is an easily accessible property in \gls{md}. The total force on each particle consists of the short-range pair forces (i.e., \gls{ikp} or Coulomb) and the long-range Lorentz force. In \cref{eq:lw}, $\bm{n}$ is the unit vector pointing from the source (at the retarded position) towards the observer, $\bm{\beta}=\bm{v}/c$ the dimensionless velocity, and $\bm{r}$ the particle position. Spectra are presented here with respect to the emission coefficient $j=2\pi(\dd^2 I/\dd \Omega/\dd \omega)/(V \tau)$, where $V$ is the domain volume and $\tau$ is the simulation time. 

\begin{figure}[!tb]
    \centering
     \includegraphics[width=\linewidth]{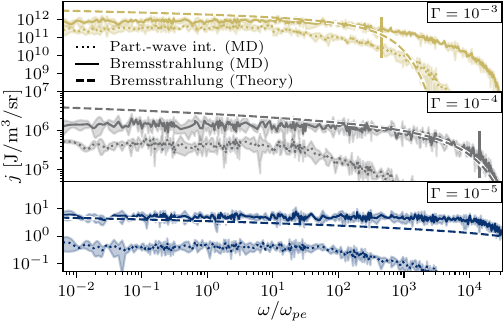}
    \caption{Radiation spectra of particle-particle and particle-wave interactions from \gls{md} compared with bremsstrahlung analytical models for $\varepsilon\approx 1$. Vertical bars denote $\omega_{phys}$.}
    \label{fig:LW_j_all}
\end{figure}

\Cref{eq:lw} is implemented in LAMMPS, where the integral is computed with the Discrete Fourier Transform method from 0 to $\tau$ integral limits. For an isotropic plasma, the emitted radiation is equivalent across all directions. All spectra were averaged over 6 observers along the $\pm (\hat{\bm{x}}, \hat{\bm{y}}, \hat{\bm{z}})$ directions of the Cartesian coordinate axes, in order to reduce numerical noise. Bremsstrahlung is computed from particle-particle forces, whereas particle-wave radiation is computed from the Lorentz forces. Species-resolved radiation spectra are computed by isolating the forces on a given species. Simulation details are discussed in \cref{sisec:sim_details}. In our simulations, the bremsstrahlung and particle-wave radiation spectra observed from different observer directions are comparable, cf. \cref{sifig:spectra_observers}, with slight deviations stemming from the finite number of particles. The possible frequency range that can be extracted from \gls{md} is $\omega \in [2\pi/\tau,\pi/\Delta t]$. In all simulations, $\tau=10^3/\omega_{pe}$ and the timestep $\Delta t=10^{-4}/\omega_{pe}$, where $\omega_{pe}=\sqrt{n_e e^2 / (m_e \epsilon_0)}$ is the plasma frequency. 

A distinction of this numerical framework compared to typical \gls{pic} approaches is that bremsstrahlung is computed directly from particle-particle forces, which are modelled explicitly and self-consistently, instead of stochastically evaluating cross-sections. Collisions also arise self-consistently in PIC, as described by the Klimontovich equation \cite{Okuda1970}, so a similar approach could in principle be applied here if each PIC particle corresponded to a physical particle. PIC resolves interactions down to the cell size of its grid, but bremsstrahlung depends on interactions down to the smallest de Broglie wavelength, $\lambda_{ei}\approx \hbar/(2\kappa T m_{e})^{1/2}$. Thus, to retain the same accuracy with our approach, the cell size in PIC would need to be $<\lambda_{ei}$, which, especially for the low-coupling simulations, would introduce a prohibitive computational cost, as detailed in \cref{app:pic_vs_md}. This is avoided in our formulation by the use of a gridless description of the short-range interactions. Note however that even then, short-range particle interactions would be determined by the numerical (spline) shape of the PIC particles, so quantum effects that we include in our framework through the \gls{ikp} would be missing. Finally, since pair and electromagnetic forces can be computed separately, the \gls{lw} spectra for each force can be obtained from the same simulation. This allows us to measure the full spectrum of synchrotron/jitter radiation.

The theoretical \gls{ei} emission is computed by integrating the \gls{bh} cross section along all possible incident electron momenta that follow a \gls{mj} distribution, while the \gls{ee} emission requires integration over both colliding electrons' momenta. The total theoretical emission is the sum of the two \gls{ee} and \gls{ei} contributions. The methodology to compute both theoretical models is discussed in Section S4, and a link to the developed code is provided. 

\begin{figure}[!tb]
    \centering
    \includegraphics[width=\linewidth]{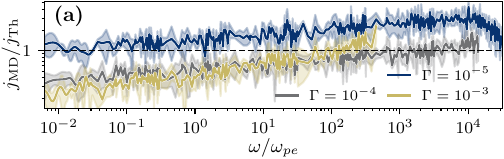}
    \includegraphics[width=\linewidth]{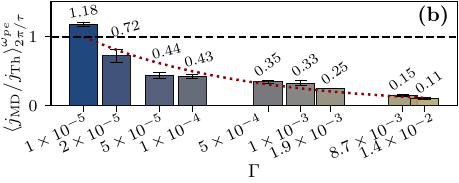}
    \caption{Bremsstrahlung \gls{md} to theory ratio at $\varepsilon\approx 1$ (a) plotted up to $\omega_{phys}$ and (b) average value for $\omega\in[2\pi/\tau,\omega_{pe}]$.}
    \label{fig:LW_j_brem_ratio}
\end{figure}

The \gls{md} emission spectra are presented in \cref{fig:LW_j_all} for different plasma conditions and are compared with the \gls{bh} and \gls{h} models. For $\Gamma=10^{-5}$, the \gls{md} bremsstrahlung agrees with the theory in the low-frequency regime, i.e., $\omega\lesssim \omega_{pe}$. The individual \gls{ee} and \gls{ei} spectra also agree with the \gls{h} and \gls{bh} models, respectively (cf. \cref{fig:LW_j_indiv}). As $\Gamma$ increases, the low-frequency \gls{md} bremsstrahlung drops below the theory, revealing the \gls{lpm} effect, which is discussed below. At high frequencies, i.e.,  $\omega\gtrsim100 \omega_{pe}$, the \gls{md} spectrum may deviate from the theoretical one for two reasons: 

First, the classical particle \gls{md} approach permits photon emission from all electrons irrespective of their energy, whereas in a quantum treatment an electron can only emit a photon with less energy compared to the electron's initial energy. This is known as the ultraviolet or Rayleigh-Jeans (RJ) catastrophe \cite{Rybicki1985,Bethe1934}. We can approximately separate this high-frequency regime in \gls{md}, by the maximum physical frequency $\omega_{phys}=\langle E\rangle/\hbar$, where the average kinetic energy of an \gls{mj} distribution is $\langle E\rangle =m_e c^2\left[K_3(1/\varepsilon)/K_2(1/\varepsilon)-\varepsilon-1\right]$ \cite{Rezzolla2013} and $K_i$ is the $i$th modified Bessel function of the second kind. $\omega_{phys}$ is denoted by a vertical bar in \cref{fig:LW_j_all}. To attenuate the RJ divergence, an ad hoc solution would be to apply a lower bound of $(\gamma-1)m_ec^2\ge\omega\hbar$ to the electron velocities in \cref{eq:lw} so that only electrons with sufficient energy emit photons \cite{Rybicki1985,Kinney2025a}, but, since the proper treatment of the high-frequency regime requires a quantum formulation, this correction is not applied here.

Second, even though the whole frequency range of the $\Gamma=10^{-5}$ spectrum is below $\omega_{phys}$, the high-frequency \gls{md} bremsstrahlung is slightly higher than the theoretical one by a factor of up to 5. Emission at this $\omega$ regime is only populated by high-velocity particles undergoing close collisions. Since the \gls{mj} distribution in \gls{md} is represented by a finite number of particles, only a few high-energy particles exist in the simulation domain, and the high-velocity distribution is not perfectly \gls{mj}. Therefore, only a limited number of such collision events exist in the \gls{md} domain, and thus the radiation spectra cannot be perfectly resolved. The high-frequency regime does not affect our conclusions since the \gls{lpm} analysis below relies on $\omega\le\omega_{pe}$, and the jitter analysis on particle-field emission at $\Gamma=10^{-5}$, which does not arise from close particle collisions and is below $\omega_{phys}$.

\Cref{fig:LW_j_brem_ratio} shows the bremsstrahlung $j$ ratio of \gls{md} to theory. For $\Gamma=10^{-5}$, the low-frequency ratio agrees within 20\% with the theory, but it decreases as $\Gamma$ increases, revealing the \gls{lpm} phenomenon. The fit to the average value inbetween the $\omega\in[2\pi/\tau,\omega_{pe}]$ limits is $\langle j_{\mathrm{MD}}/j_{\mathrm{Th}} \rangle \approx A(10^4 \Gamma)^b$, where $\Gamma\ge 10^{-5}$, $A=0.5$, and $b=-0.3$. Even when $\varepsilon=0.1$, simulations at $\Gamma\ge10^{-4}$ showed that the \gls{lpm} effect is significant (cf. \cref{sifig:LW_j_brem_ratio_bar_lowT}). Therefore, even within the weak coupling regime ($\Gamma > 10^{-5}$), \gls{lpm} still substantially affects bremsstrahlung. This indicates that for our electron-proton plasma, a $\Gamma \le 10^{-5}$ is needed to justify the application of the \gls{bh} and \gls{h} models. The current \gls{md} approach can be used to investigate the impact of \gls{lpm} in different plasma conditions and thus test the application of the \gls{bh} and \gls{h} cross sections in \gls{pic} simulations.

Typically, simulating plasma phenomena necessitates resolving the Debye length $\lambda_D$. In many simulations presented in \cref{fig:LW_j_brem_ratio}b, the domain length $L$ is $<\lambda_D$ (cf. \cref{sisec:sim_details}). Resolving $\lambda_D$ at low $\Gamma$ values requires a huge particle number, leading to a computationally prohibitive cost. Nonetheless, resolving $\lambda_D$ does not affect bremsstrahlung, as evident in  \cref{sifig:LW_j_all_Debye}. Specifically, comparing bremsstrahlung without and with ($L/\lambda_D=5$) resolution of $\lambda_D$ showed no difference. This is explained because bremsstrahlung stems from isolated particle interactions at the sub-Debye scale. The particle-wave radiation differs between the two domains, which is expected, as the smaller domain, with only 500 particles, does not contain enough $\bm{k}$-modes to resolve the field fluctuations fully. Our main simulations, however, with $10^4$ particles, contain enough modes, since, as shown in \cref{sifig:LW_j_sim_15_better}, increasing the number of particles to $5\times10^4$ does not change the particle-wave radiation. The study of more collective phenomena that require a plasma (e.g., Ref. \cite{Tsytovich1996}) would need $L\gg\lambda_D$ and is left for future work.

\begin{figure}
    \centering
    \includegraphics[width=0.5\linewidth]{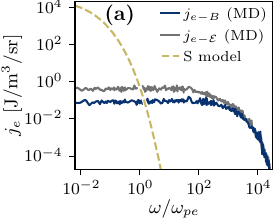}%
    \includegraphics[width= 0.5\linewidth]{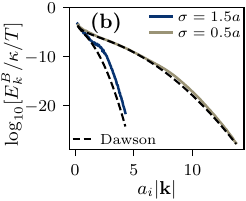}

    \includegraphics[width=0.5\linewidth]{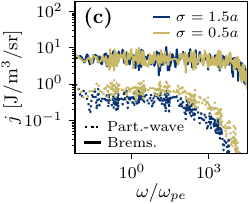}%
    \includegraphics[width=0.5\linewidth]{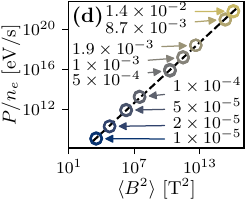}

    \caption{(a) $e$-$B$ and $e$-$\mathcal{E}$ radiation spectra at $\Gamma=10^{-5}$, $\varepsilon\approx1$ compared with the S analytical model; (b) $E_k^B$ spectra comparison between MD and \cref{eq:Dawson}; (c) MD bremsstrahlung and particle-wave radiation spectra for two $\sigma$ at $\Gamma=10^{-5}$ and $\varepsilon\approx1$; and (d) total $B$ power at different $\Gamma$.}
    \label{fig:LW_j_eB}
\end{figure}

\Cref{fig:LW_j_all} shows that particle-wave interactions become significant at low frequencies compared to bremsstrahlung especially as $\Gamma$ increases. At $\Gamma=10^{-5}$ particle-wave radiation reaches 6 times smaller than bremsstrahlung at $\omega\approx\omega_{pe}$, whereas at $\Gamma=10^{-3}$ both $j$ values become comparable. The particle-wave radiation originates from electron-magnetic-field and electron-electric-field interactions, as is evident from \cref{fig:LW_j_eB}a. This is expected since the induced electromagnetic waves have $|\bm{\mathcal{E}}|\approx c|\bm{B}|$, and thus the electric to the magnetic Lorentz force components differ by $\approx\beta^{-1}$, which, for highly relativistic electrons, tends to 1. Below, we focus on radiation induced by magnetic field fluctuations, which, unlike the electric field ones, are not screened at the Debye length \cite{Bellan2008}. Note also that the electrostatic approximation of the short-range interactions acts as a filter that dissipates the energy of the high $\bm{k}$ modes, effectively shielding the \gls{md} simulations from the RJ catastrophe, characteristic in \gls{pic} simulations \cite{More2010}. Moreover, radiation reaction is negligible in our simulations (cf. Section S6.1).

In a thermalized particle simulation, the spectrum of the average magnetic energy according to Dawson is \cite{Dawson1983}
\begin{equation}
\langle E_k^B \rangle_T = \frac{\kappa T}{2 + 2\lambda_{\mathrm{EM}}^2 \bm{k}^2 e^{\sigma^2 \bm{k}^2}},
\label{eq:Dawson}
\end{equation}
where $\lambda_{\mathrm{EM}} = \lambda_D\sqrt{\langle \gamma \rangle_T / \varepsilon}$ is the skin depth with the relativistic correction and $\langle \gamma \rangle_T$ is computed by averaging over the \gls{mj} distribution. As shown in \cref{fig:LW_j_eB}b, the spectrum from \gls{md} simulations matches the Dawson spectrum at low $k$. Decreasing $\sigma$ improves the $E_k^B$ agreement, because more $\bm{k}$-modes are included and short-range contributions, which scale as $\text{erfc}[r_c/(2\sigma)]$, where $r_c$ is the short-range interaction cutoff (cf. Section S3), are reduced. Rerunning the simulations with $\sigma = 0.5 a$ reduces this factor by $\sim10$ orders of magnitude, and results in a good agreement of $E_k^B$ between the MD and theory. However, this decrease in the Gaussian widths increases the required $k$-space grid resolution and more than doubles the computational cost. Nonetheless, the \gls{lw} spectrum is not considerably affected by the $\sigma$ decrease, as shown in \cref{fig:LW_j_eB}c. Specifically, the ratio of the total powers, $P_{\sigma = 0.5 a}/P_{\sigma = 1.5 a}$, where $P=4\pi\int j \dd \omega$, is equal to 1.0 for bremsstrahlung and 2.3 for particle-wave radiation.

The high-frequency radiation shown in $j_{e-B}$ of \cref{fig:LW_j_eB}a is attributed to jitter radiation. As shown by Medvedev \cite{Medvedev2000}, jitter radiation affects primarily the high-frequency regime and thus we expect that magnetic field fluctuations lead to the observed $j_{e-B}$ spectrum. To the best of our knowledge, a jitter radiation model for weakly relativistic electrons in an isotropic plasma with wavelike magnetic perturbations that match the Dawson model does not exist in the literature. First, we compute the relation of $P_{e-B}$ to the average magnetic field strength $\langle B^2 \rangle_T=\sum_{|\bm{k}|}\langle B^2 (\bm{k}) \rangle_T$. In $e$-$B$ radiation, $P\propto n_e B^2$ \cite{Landau1975}, which, as shown in \cref{fig:LW_j_eB}d, is the exact ratio observed in our \gls{md} simulations. Next, we compute the analytical synchrotron radiation emitted from an accelerated electron moving in a perfectly circular motion from the \gls{s} model \cite{Schwinger1949}, for the magnetic fields extracted from the \gls{md} simulations, as detailed in \cref{app:sync_model}. The comparison for $\Gamma=10^{-5}$ is shown in \cref{fig:LW_j_eB}a. The power, $P$, of the MD spectrum agrees within 39\% with the power of the \gls{s} model. This agreement is expected because synchrotron and jitter radiation emit the same total power and only differ in the frequency distribution of that power \cite{Landau1975}. Thus, the electrons emit the correct power for the measured magnetic field, but at the higher frequencies of jitter radiation. Finally, for jitter radiation, the magnetic field modes must satisfy $\zeta<1$, where $\zeta(k)=2\pi e B(k)/(m_e c k)$ is the strength parameter \cite{Reville2010}. In all simulations, $\zeta<1 \forall k$, suggesting that the particles radiate in the jitter regime. These findings confirm the significant impact of jitter radiation in our electron-proton relativistic plasmas, and indicate that it cannot be neglected at the studied conditions.

The photon formation length $\ell_f(\omega)$ \cite{Akhiezer,Reville2010} is another important length scale and corresponds to the distance the electron traverses before the emitted photon is fully formed. It is computed from $\ell_f(\omega)\approx c \tau_f(\omega)$ for highly relativistic electrons, where the formation time period at each frequency is found from $\tau_f - |\bm{r} (t + \tau_f) - \bm{r} (t)|/c = 2 \pi/\omega$, taking $\bm{n}\parallel\bm{\beta}$. If $L<\ell_f(\omega)$, the domain is too small to capture the full electron trajectory and corresponding electromagnetic fields during the photon formation process, which affects the low-frequency radiation. At large $\Gamma$ values, our domain length is $<\ell_f$ (cf. Fig.~S7). Resolving $\ell_f$ is necessary for jitter radiation, because a domain with $L<\ell_f$ cannot contain the low $\bm{k}$-modes that contribute to the low-frequency jitter radiation, thus introducing a low-$\bm{k}$ cutoff. This is not an issue for bremsstrahlung, stemming from binary collisions, which are well resolved if $L\gg a$. To validate this, we compare the spectra from simulations with different domain sizes. As shown in \cref{sifig:LW_j_sim_15_better}, we find no impact on bremsstrahlung, while negligible low-frequency differences are found in particle-wave radiation. The fact that $\ell_f$ is not resolved at low frequencies does not affect the high-frequency jitter emission, which remains well resolved. Also, if $c \Delta t>\ell_f(\omega)$, then the electron path is not properly resolved, leading to inaccurate large-frequency radiation measurements. Our simulations have a sufficiently small $\Delta t$ for all cases (cf. Fig.~S7).

In conclusion, this study presents a novel \gls{md} framework to investigate radiation processes in relativistic plasmas. Results show that at coupling constants above $10^{-5}$ the \gls{lpm} phenomenon becomes significant, leading to a decrease in bremsstrahlung. At $\Gamma\approx 10^{-2}$, this decrease is 90\% of the theoretical value. These results serve as a guideline for when \gls{pic} simulations of relativistic solid-density plasmas can neglect the \gls{lpm} effect. Additionally, results show that electron-magnetic field radiation, whose high-$k$ contribution is typically not captured in \gls{pic} simulations, becomes comparable in magnitude to bremsstrahlung at higher coupling. Overall, this new framework enables computation of radiation from first principles and will pave the way for the development of new bremsstrahlung suppression and magnetic field-induced radiation models.

\bigskip

\section*{Data and Code Availability}
The analytical bremsstrahlung code (\texttt{pyRelBrem}) used to compute the emission spectra is publicly available on GitHub at \url{https://github.com/ekritikos/pyRelBrem}.

\begin{acknowledgments}
This work was funded in part by the UK’s Department for Energy Security and Net Zero as part of UPLiFT, the ``UK Programme of Laser Inertial Fusion Technology for Energy.'' PS acknowledges funding from the Fusion2024 program of the German Federal Ministry of Research, Technology and Space (BMFTR) via the project ``VANLIFE'' (funding no 13F1016B). This work used the ARCHER2 UK National Supercomputing Service (https://www.archer2.ac.uk).
\end{acknowledgments}

\bibliography{references}

\clearpage
\onecolumngrid
\begin{center}
\textbf{\large End Matter}
\end{center}
\twocolumngrid
\raggedbottom

\appendix
\setcounter{secnumdepth}{1}  
\crefalias{section}{appendix}

\section{PIC vs MD capabilities}
\label{app:pic_vs_md}
\vspace{-10pt}

PIC simulations can in principle describe the same kinetics as our MD simulations; however, this would lead to a reduction in accuracy and significant increase in computational cost. First, our MD approach describes particle interactions explicitly at all separations, i.e., classically at separations larger than the de Broglie wavelength $\lambda$, and including quantum diffraction effects at separation distances smaller than $\lambda$ via the \gls{ikp} \cite{Filinov2004}. In contrast, close particle interactions in PIC simulations are smoothed by the numerical shape of the modelled particles \cite{Arber2015}. Therefore, PIC simulations miss the interaction at separations below the cell size and quantum effects that are included in our MD framework. 

Even if quantum pair-corrections are included, PIC can resolve these interactions only if the grid spacing is $\Delta x<\lambda_{ij}$. At the current simulation conditions, where $\varepsilon=1 \Rightarrow T= m_e c^2/\kappa$, the de Broglie wavelengths for the two species interactions are: (i) $ \lambda_{ee} = \hbar/(\kappa T m_{e})^{1/2}=3.9\times 10^{-3}$ {\AA} and (ii) $\lambda_{ei} \approx \hbar/(2\kappa T m_{e})^{1/2}=2.7\times 10^{-3}$ {\AA}, i.e., $\Delta x<\min \left(\lambda_{ee},\lambda_{ei}\right)=\lambda_{ei}$. Our MD simulations also require a grid for the long-range electromagnetic solver; however, since close-range particle interactions are computed directly, the grid spacing is not constrained by them and thus $\Delta x\approx a$. Therefore, the ratio of the PIC cells, $N_{\rm PIC}$ to the number of MD cells, $N_{\rm MD}$ is
\begin{equation}
    \dfrac{N_{\rm PIC}}{N_{\rm MD}}=\dfrac{(\Delta x_{\rm MD})^3}{(\Delta x_{\rm PIC})^3}>  \left(\frac{a}{\lambda_{ei}}\right)^{3}=\dfrac{3(\kappa T m_{e})^{3/2}}{2^{1/2} \pi n \hbar^3}.
    \end{equation}
Moreover, $a/\lambda_{ei}=2^{1/2}\alpha_f/(\varepsilon^{1/2}\Gamma)$ where $\alpha_f$ is the fine-structure constant. Therefore, to keep the MD level of resolution, a PIC simulation would require $>10^9$ times more cells than our MD simulations at $\Gamma=10^{-5}$, and more than $10^{3}$ times more cells at $\Gamma=10^{-3}$.

\vspace{-10pt}
\section{Improved Kelbg Potential Details}
\label{app:ikp}
\vspace{-10pt}

The pair potential energy of the \gls{ikp} model is \cite{Filinov2004}
\begin{equation}
U_{i j} = \frac{k_c q_i q_j}{r_{i j}} \left\{ 1 - e^{- \left( \frac{r_{i j}}{\lambda_{ij}} \right)^2} + \frac{r_{i j}\sqrt{\pi}}{\lambda_{ij} \chi } \left[ 1 -\erf \left( \frac{\chi r_{i j}}{\lambda_{ij}} \right) \right] \right\},
\label{eq:ikp_pot}
\end{equation}
where for \gls{ei} interactions $\chi = (x + x^2)/(1 + a_{e p} x + x^2)$, with $x = \sqrt{8 \pi T / 315775}$ and $a_{e p} = 1.09$. For \gls{ee} interactions $\chi(T)=[\chi(T\rightarrow0)+a_{ee}x+x^2]/(1+x^2)$, where $a_{ee}=0.18$, and $\chi(T\rightarrow0) \approx- 2\tilde{x}^3/\{\sqrt{\pi} [\ln(8\tilde{x}^4/\sqrt{\pi})-3\tilde{x}^2)]\}$, with $\tilde{x}=|\pi e^2/(8\pi \epsilon_0 \kappa T \lambda_{ee})|^{1/3}$. The \gls{ikp} model is plotted in Fig.~S1 for \gls{ei} interactions.

\vspace{-10pt}
\section{Simulation details}
\label{sisec:sim_details}
\vspace{-10pt}

All simulations are performed with cubic domains and periodic boundary conditions. The simulation conditions of all cases are shown in \cref{tab:simulation_conditions}. In \cref{tab:simulation_conditions}, $\Theta=\kappa T/E_F$ is the degeneracy parameter and $E_F=\hbar^2(3\pi^2n_e)^{2/3}/(2 m_e)$ is the Fermi energy of a free electron gas. Additionally, $N_{sim}$ denotes the number of simulations performed for the respective conditions. Repeated simulations differ in the initial particle position and velocity distributions. Overall, we investigate plasmas of $\Gamma\in[10^{-5},10^{-2}]$ and $\varepsilon\in[0.1,1]$. The pair potential cutoff $r_c$ is set to $5a$. The timestep used here is sufficiently small to resolve both the $\sigma$ and $a$ length scales, i.e., $\dd t\ll \sigma/c$ and $\dd t\ll a/c$.

Initially, particles are randomly distributed in the domain and undergo an energy minimization step. Next, random initial velocities are assigned to each particle sampled from an MJ distribution. The net momentum of the particles is adjusted to zero. The particles are then energy equilibrated for a total simulation time of $10^2/\omega_{pe}$, where the temperature of the system is rescaled every 500 timesteps to the target temperature. Finally, the data collection is performed without rescaling. 

The radiation spectra are computed with the \gls{lw} fields at every timestep. For most conditions (cf. \cref{tab:simulation_conditions}), 3 simulations are performed, differing in the initial position and velocity distributions. The presented spectra are averaged over the 3 simulations. The spectra are also averaged over the different observers (equal to 6 unless otherwise specified in the main paper), located in the directions $\pm \hat{\bm{x}}$, $\pm \hat{\bm{y}}$, and $\pm \hat{\bm{z}}$ of the Cartesian coordinate axes. It is assumed that the observer is at infinity, and so the $\bm{n}$ vector is always parallel to the coordinate axes.

\vspace{-10pt}
\section{Synchrotron model}
\label{app:sync_model}
\vspace{-10pt}

The analytical synchrotron radiation emitted from an accelerated electron moving in a perfectly circular motion is given by the \gls{s} model \cite{Schwinger1949}, which is detailed in Section S5. We evaluate the \gls{s} model with the magnetic field measured in the \gls{md} domains. For our isotropic plasma, we compute the magnetic field strength $\langle B^2 (k) \rangle_T^{1/2}$ for each wavenumber $k=|\bm{k}|$, where $\langle B^2 (k) \rangle_T=2\mu_0 \langle E^B (k) \rangle_T/V$ and $E^B (k)$ is the total magnetic energy for each $k$, averaged over time and over the three $\Gamma = 10^{-5}$ simulations. When computing the \gls{s} spectrum, we accumulate the individual radiation spectra computed from the 40 smallest $k$. At larger $k$ values, the magnetic energy decreases exponentially and thus does not contribute to the overall spectrum (the first 40 $k$ numbers hold $>98$\% of the total $E^B$). As shown in \cref{fig:LW_j_eB}a, the analytical synchrotron emission is higher than the \gls{md} spectrum at low frequencies and quickly decays as $\omega \rightarrow \omega_{pe}$ for our conditions. This decay occurs because the electron motion is considered periodic and so the total spectrum is a combination of harmonics of the rotational angular frequency $\omega_0=v/r_l$, where $r_l=\beta \gamma m_e c/(e B)$ is the Larmor radius of the trajectory. Therefore, for the same $B$ strength, increasingly high-velocity electrons are needed to populate the high-frequency regime, but the \gls{mj} distribution decreases as $v \rightarrow c$.

\onecolumngrid
\begin{center}
\begin{minipage}{\textwidth}
\vspace{-4pt}
\captionof{table}{Simulation conditions.}
\label{tab:simulation_conditions}
    \centering
\begin{tabular*}{\textwidth}{@{\extracolsep{\fill}} ccccccccccccc @{}}
      \hline
      No. & $\Gamma$ & $\Theta$ & $\varepsilon$ & $T$ [keV] & $n$
      [m\textsuperscript{$- 3$}] & $N_{\ensuremath{\operatorname{all}}}$ & $L$
      [\AA] & $a$ [\AA] & $L / \lambda_D$ & $N_{sim}$ & CPUs/run & CPU-hours/run\\
      \hline
      1 & $10^{- 5}$ & 468943 & 0.978 & 500 & $10^{28}$ & $10^4$ & 100 & 2.879 & 0.13 & 3 & 128 & 9580$^\dagger$\\
      2$^*$ & $10^{- 5}$ & 468943 & 0.978 & 500 & $10^{28}$ & $5 \times 10^4$ & 171 & 2.879 & 0.23 & 1 & 384 & 22821\\
      3 & $2 \times 10^{- 5}$ & 117236 & 0.978 & 500 & $8 \times 10^{28}$ & $10^4$ & 50 & 1.440 & 0.19 & 3 & 128 & 4795\\
      4 & $5 \times 10^{- 5}$ & 18757.7 & 0.978 & 500 & $1.25 \times 10^{30}$ & $10^4$ & 20 & 0.576 & 0.30 & 3 & 128 & 4811\\
      5 & $10^{- 4}$ & 4689.43 & 0.978 & 500 & $10^{31}$ & $10^4$ & 10 & 0.288 & 0.43 & 3 & 128 & 9557$^\dagger$\\
      6 & $5 \times 10^{- 4}$ & 187.58 & 0.978 & 500 & $1.25 \times 10^{33}$ & $10^4$ & 2 & 0.058 & 0.95 & 3 & 128 & 4863\\
      7 & $10^{- 3}$ & 46.89 & 0.978 & 500 & $10^{34}$ & $10^4$ & 1 & 0.029 & 1.35 & 3 & 128 & 9745$^\dagger$\\
      8 & $1.9 \times 10^{- 3}$ & 13.34 & 0.978 & 500 & $6.59 \times 10^{34}$ & $2 \times 10^5$ & 1.45 & 0.015 & 5.0 & 1 & 1280 & 102044 \\
      9$^*$ & $1.9 \times 10^{- 3}$ & 13.34 & 0.978 & 500 & $6.59 \times 10^{34}$ & 500 & 0.20 & 0.015 & 0.96 & 3 & 3 & 32 \\
      10 & $8.7 \times 10^{- 3}$ & 0.62 & 0.978 & 500 & $6.59 \times 10^{36}$ & $2 \times 10^4$ & $0.14$ & 0.003 & 5.0 & 3 & 256 & 10202\\
      11 & $1.4 \times 10^{- 2}$ & 0.25 & 0.978 & 500 & $2.64 \times 10^{37}$ & $10^4$ & 0.07 & 0.002 & 5.0 & 3 & 128 & 4711 \\
      12 & $10^{- 4}$ & 46894.3 & 0.098 & 50 & $10^{28}$ & $10^4$ & 100 & 2.879 & 0.43 & 3 & 128 & 4774\\
      13 & $10^{- 3}$ & 468.94 & 0.098 & 50 & $10^{31}$ & $10^4$ & 10 & 0.288 & 1.35 & 3 & 128 & 4814\\
      14 & $10^{- 2}$ & 4.69 & 0.098 & 50 & $10^{34}$ & $10^4$ & 1 & 0.029 & 4.25 & 3 & 128 & 4679\\
      \hline
\multicolumn{13}{l}{$^*$Supplementary simulations performed to compare with the main runs; explicitly specified when used in the text.} \\    
\multicolumn{13}{l}{$^\dagger$This run time includes the additional \gls{lw} spectra computed separately for the $e$-$B$ and $e$-$\mathcal{E}$ interactions.} \\
\end{tabular*}
\end{minipage}
\end{center}
\twocolumngrid

\section{Additional simulation results}
\label{app:add_results}

\noindent\begin{minipage}{\linewidth}
    \centering
    \includegraphics[width=0.9\linewidth]{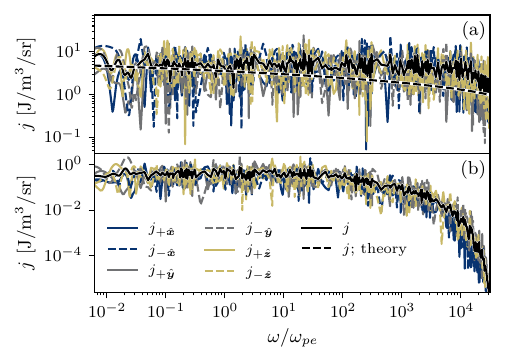}
    \captionof{figure}{(a) Bremsstrahlung and (b) particle-wave radiation spectra computed at different observer directions from the simulation case with $\Gamma = 10^{-5}$ and $\varepsilon\approx 1$.}
    \label{sifig:spectra_observers}
\end{minipage}\par\bigskip

\noindent\begin{minipage}{\linewidth}
    \centering
    \includegraphics[width=0.9\linewidth]{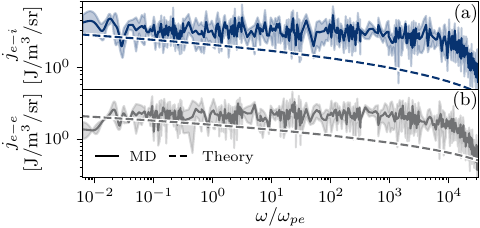}
    \captionof{figure}{\gls{md} bremsstrahlung spectra from (a) $e$-$i$ and (b) $e$-$e$ interactions compared with the \gls{bh} and \gls{h} analytical models (theory) for $\Gamma=10^{-5}$ and $\varepsilon\approx1$. The filled area shows the standard deviation.}
    \label{fig:LW_j_indiv}
\end{minipage}\par\bigskip

\noindent\begin{minipage}{\linewidth}
    \centering
    \includegraphics[width=0.87\linewidth]{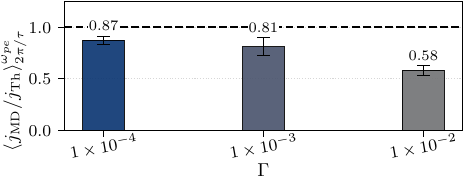}
    \captionof{figure}{Bremsstrahlung \gls{md} to theory ratio averaged for $\omega\in[2\pi/\tau,1]\omega_{pe}$ for the plasma case with $\varepsilon\approx 0.1$.}
    \label{sifig:LW_j_brem_ratio_bar_lowT}
\end{minipage}\par\bigskip

\noindent\begin{minipage}{\linewidth}
    \centering
    \includegraphics[width=0.9\linewidth]{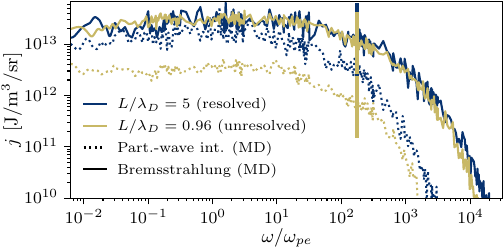}
    \captionof{figure}{Radiation spectra measured from simulations with ($L/\lambda_D=5$) and without ($L/\lambda_D=0.96$) sufficient resolution of $\lambda_D$ at $\Gamma = 1.9\times 10^{-3}$ and $\varepsilon\approx 1$ corresponding to the simulation case numbers (cf. \cref{tab:simulation_conditions}) of 8 and 9.}
    \label{sifig:LW_j_all_Debye}
\end{minipage}\par\bigskip

\noindent\begin{minipage}{\linewidth}
    \centering
    \includegraphics[width=0.9\linewidth]{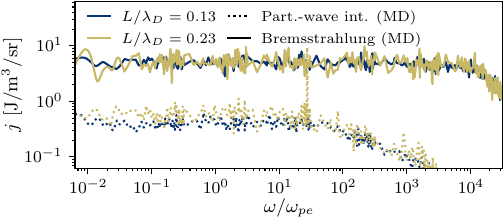}
    \captionof{figure}{Radiation spectra measured from simulations with different domain lengths (cases with no. 1 and 2 in \cref{tab:simulation_conditions}) at $\Gamma = 10^{-5}$ and $\varepsilon\approx 1$.}
    \label{sifig:LW_j_sim_15_better}
\end{minipage}\par\bigskip

\end{document}


\title{\textbf{Radiation processes in relativistic plasmas via particle simulations}}

\author{Efstratios M. Kritikos}
\email{efstratios.kritikos@physics.ox.ac.uk}

\affiliation{Department of Physics, University of Oxford, Parks Road, Oxford OX1 3PU, United Kingdom}

\author{Pontus Svensson}

\affiliation{Institute of Radiation Physics, Helmholtz-Zentrum Dresden-Rossendorf (HZDR), D-01328 Dresden, Germany}

\author{Daniel Plummer}

\affiliation{Department of Physics, University of Oxford, Parks Road, Oxford OX1 3PU, United Kingdom}

\author{Brian Reville}

\affiliation{Max-Planck-Institut für Kernphysik (MPIK), Saupfercheckweg, Heidelberg D-69117, Germany}

\author{Bob Bingham}

\affiliation{Central Laser Facility, STFC Rutherford Appleton Laboratory, Didcot, OX11 0QX, United Kingdom}

\author{Gianluca Gregori}

\affiliation{Department of Physics, University of Oxford, Parks Road, Oxford OX1 3PU, United Kingdom}

\date{August 25, 2026}

\begin{abstract}
    This document reports additional information in support of the discussion of the results presented in the main paper. The structure of the Supporting Information is as follows:
\end{abstract}

\maketitle

\tableofcontents

\section{Improved Kelbg Potential}
\label{sisec:ikp}

\begin{figure}[!tb]
    \centering
    \includegraphics[width=0.5\linewidth]{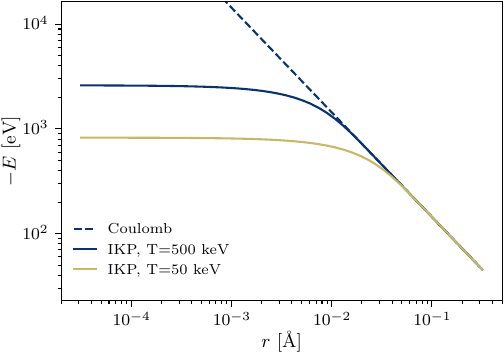}%
    \includegraphics[width=0.5\linewidth]{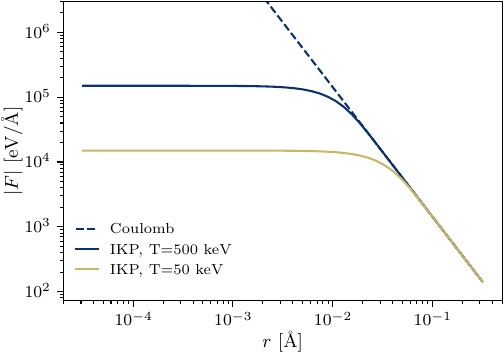}
    \caption{Comparison of the potential energy (left) and force (right) of the \gls{ikp} and Coulomb models for an electron-proton interaction at different temperatures.}
    \label{sifig:IKP_th}
\end{figure}

\begin{figure}
    \centering
    \includegraphics[width=0.525\linewidth]{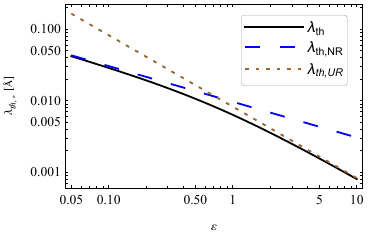}%
    \includegraphics[width=0.475\linewidth]{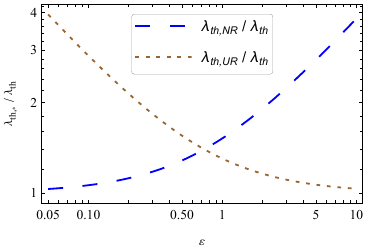}
    \caption{Plot of $\lambda_{th, NR}$, $\lambda_{th, UR}$, and $\lambda_{th}$ (left); and of the $\lambda_{th, NR}/\lambda_{th}$, $\lambda_{th, UR}/\lambda_{th}$ ratios (right).}
    \label{sifig:lambda_rel}
\end{figure}

The \gls{ikp} for the different conditions studied in this work are plotted in \cref{sifig:IKP_th}. As mentioned in the main paper, a possible way to apply relativistic effects in the \gls{ikp} is to include the relativistic de Broglie wavelength. The full form of the thermal wavelength is \cite{Hao2023}
\begin{equation}
  \lambda_{th} = \frac{\lambda_C}{[4 \pi e^{1/\varepsilon} \varepsilon K_2 (1/\varepsilon)]^{1/3}},
\end{equation}
where $\lambda_C = h / (m c)$ is the Compton wavelength, $K_2$ is the modified Bessel function of the second kind, and $\varepsilon = \kappa T/m c^2$. From this equation we can extract two extreme cases, i.e., the nonrelativistic case (NR for which $\varepsilon \ll 1$) and the ultrarelativistic (UR for which $\varepsilon \gg 1$). The respective equations for these two cases are
\begin{equation}
  \lambda_{th, NR} = \frac{h}{\sqrt{2 \pi m \kappa T}},
\end{equation}
and
\begin{equation}
  \lambda_{th, UR} = \frac{c h}{2 \pi^{1 / 3} \kappa T}.
\end{equation}
The ratio of $\lambda_{th, NR}$ and $\lambda_{th, UR}$ to the $\lambda_{th}$ are plotted in \cref{sifig:lambda_rel}. It is evident that for $\varepsilon\le1$, the $\lambda_{th, NR}$ is a good approximation to $\lambda_{th}$. Therefore, the $\lambda_{th, NR}$ has been used for the numerical computations in this work.

\section{Code details}
\label{sisec:code_details}

Simulations are performed in the code developed in Ref. \cite{Svensson2024}. The framework works by separating the long-range and short-range interactions. Formally, this is done by splitting the charge and current densities into two components. First, a smooth description of the charge and current densities, and second, the highly fluctuating difference between the smooth and the exact densities. The smoothed charge and current densities and the corresponding electromagnetic fields can be efficiently treated in Fourier space and using grid methods, where the electromagnetic fields are time-evolved to describe retardation and electromagnetic interactions. The remaining charge and current density profiles are, by construction, charge neutral on length scales large compared to $\sigma$, and most interactions are screened on length scales large compared to $\sigma$. This makes it suitable to evaluate using real space techniques in MD, here performed in the electrostatic approximation. Formally, the methodology is based on performing this split in the Lagrangian for the particle-field system, as described in Ref. \cite{Svensson2024}, resulting in self-consistent dynamics. Heuristically, the long-range interactions should account for all field effects in the plasma, whereas the short-range interaction is our approximation for the collisional behaviour. 

Practically, the smoothed charge and current densities are constructed by assuming Gaussian-distributed particles with a standard deviation $\sigma$. These densities are assigned on a grid by a 5th-order spline shape function, and by the appropriate influence functions \cite{Hockney2021}, the force on the particles and current for the time evolution of the fields are evaluated, similar to the PPPM method. Time evolution of the electromagnetic fields is carried out in the Fourier domain.

\section{Simulated electromagnetic fields}
\label{sisec:sim_elec_fields}

Within the relativistic code, the energy of the transverse electric and magnetic fields per $\bm{k}$ mode and polarization $\bm{\epsilon}^{(i)}$ is \cite{Svensson2024}
\begin{equation}
E_k^{\mathcal{E} (i)} = \frac{1}{2 \varepsilon_0 V}
   (\bm{\Pi}_{\bm{k}}^{(l)} \cdot
   \bm{\epsilon}^{(i)}) (\bm{\Pi}_{-
   \bm{k}}^{(l)} \cdot
   \bm{\epsilon}^{(i)}), 
\end{equation}
and
\begin{equation}
E_k^{B (i)} = \frac{| \bm{k} |^2}{2 \mu_0 V}
   (\bm{A}_{\bm{k}}^{(l)} \cdot
   \bm{\epsilon}^{(i)}) (\bm{A}_{-
   \bm{k}}^{(l)} \cdot
   \bm{\epsilon}^{(i)}),
\end{equation}
respectively, where the superscript $(l)$ denotes the long-range PPPM solver of the EM fields, $\bm{A}_{\bm{k}}^{(l)}$ is the magnetic vector potential and $\bm{\Pi}_{-\bm{k}}^{(l)}=\varepsilon_0 \partial \bm{A}_{\bm{k}}^{(l)}/\partial t$. The negative wave vector $-\bm{k}$ is the complex conjugate of $\bm{k}$. There are two independent polarizations per $\bm{k}$ mode, $ \bm{\epsilon}^{(1)}$ and $\bm{\epsilon}^{(2)}$, perpendicular to $\bm{k}$. 

The code outputs the real and imaginary parts of the above vectors along the two polarizations. Therefore, the energies are computed directly from the equations (dropping the $(l)$ superscripts) as
\begin{equation}
    E_k^{\mathcal{E} (i)} = \frac{1}{2 \varepsilon_0}
   [\bm{\Pi}_{\bm{k}}^{(i)}
   (\bm{\Pi}_{\bm{k}}^{(i)})^{\mathord{*}}]
   = \frac{1}{2 \varepsilon_0} \{ (\mathrm{Re}
   [\bm{\Pi}_{\bm{k}}^{(i)}])^2
   + (\mathrm{Im}
   [\bm{\Pi}_{\bm{k}}^{(i)}])^2
   \}
\end{equation}
and
\begin{equation}
   E_k^{B (i)} = \frac{| \bm{k} |^2}{2 \mu_0}
   [\bm{A}_{\bm{k}}^{(i)}
   (A_{\bm{k}}^{(i)})^{\mathord{*}}] = \frac{|
   \bm{k} |^2}{2 \mu_0} \{
   (\mathrm{Re}
   [\bm{A}_{\bm{k}}^{(i)}])^2
   + (\mathrm{Im}   [A_{\bm{k}}^{(i)}])^2 \} .
\end{equation}
The total electric field energy is $E_k^{\mathcal{E}} = E_k^{\mathcal{E} (1)} + E_k^{\mathcal{E} (2)}$ and the total magnetic field energy is $E_k^B = E_k^{B (1)} + E_k^{B (2)}$.

The average magnetic energy according to Dawson is given by Eq.~(3). The average $\gamma$ function in the $\lambda_{EM}$ term is given by
\begin{equation}
\langle \gamma \rangle_T =
   \frac{\int_0^{\infty} \sqrt{t} (1 + \varepsilon t)^2 (1 + \varepsilon t /
   2)^{1 / 2} e^{- t} \dd t}{\int_0^{\infty} \sqrt{t} (1 + \varepsilon t) (1 +
   \varepsilon t / 2)^{1 / 2} e^{- t} \dd t} .
\end{equation}

Additionally, in the Coulomb gauge, the Coulomb pair potential between particles $i$ and $j$ is 
\begin{equation}
    \phi_C^{(s)} = \frac{1}{4\pi\epsilon_0} \frac{q_i q_j}{|\bm{r}_i-\bm{r}_j|}\text{erfc}\left(\frac{|\bm{r}_i-\bm{r}_j| }{2\sigma}\right),
\end{equation}
while the IKP pair potential is
\begin{equation}
    \phi_{IKP}^{(s)} = \frac{1}{4\pi\epsilon_0} \frac{q_i q_j}{|\bm{r}_i-\bm{r}_j|}\left\{ \text{erfc}\left(\frac{|\bm{r}_i-\bm{r}_j|} {2\sigma}\right) - e^{- \left( \frac{|\bm{r}_i-\bm{r}_j|}{\lambda} \right)^2} + \frac{|\bm{r}_i-\bm{r}_j|\sqrt{\pi}}{\lambda \chi } \left[ 1 -\erf \left( \frac{\chi |\bm{r}_i-\bm{r}_j|}{\lambda} \right) \right] \right\}.
\end{equation}

Thus, sufficient screening of the interaction at the cutoff, i.e., for $|\bm{r}_i-\bm{r}_j|\rightarrow r_c$, requires $\text{erfc}[r_c/(2\sigma)]\ll 1$. The reader is reminded that the \gls{ikp} converges to the Coulomb potential near $r_c$.

\section{Theoretical relativistic bremsstrahlung}
\label{sisec:th_brem}

\subsection{Electron-ion bremsstrahlung}

According to the \gls{bh} model \cite{Bethe1934}, the cross section of an electron-ion interaction (assuming that the ion is stationary) at a given frequency $\nu$ is
\begin{equation}
\begin{aligned}
  \Phi = &\frac{Z^2}{137} r_e^2 \frac{\tilde{p}}{\tilde{p}_0} \frac{d k}{k} \left\{ \frac{4}{3}
  - 2 \tilde{E}_0 \tilde{E} \frac{\tilde{p}^2 + \tilde{p}_0^2}{\tilde{p}^2 \tilde{p}_0^2} + \mu^2 \left( \frac{\epsilon_0
  \tilde{E}}{\tilde{p}_0^3} + \frac{\epsilon \tilde{E}_0}{\tilde{p}^3} - \frac{\epsilon \epsilon_0}{\tilde{p}
  \tilde{p}_0} \right) \right.  \\
  &+ \left[\frac{8}{3} \frac{\tilde{E}_0 \tilde{E}}{\tilde{p}_0 \tilde{p}} + \frac{k^2}{\tilde{p}_0^3 \tilde{p}^3} (\tilde{E}_0^2 \tilde{E}^2
   + \tilde{p}_0^2 \tilde{p}^2) \right] L \\
  & + \frac{\mu^2 k}{2 \tilde{p} \tilde{p}_0} \left[ \frac{\tilde{E}_0 \tilde{E} +
   \tilde{p}_0^2}{\tilde{p}_0^3} \epsilon_0 - \frac{\tilde{E}_0 \tilde{E} + \tilde{p}^2}{\tilde{p}^3} \epsilon + \frac{2
   k \tilde{E}_0 \tilde{E}}{\tilde{p}^2 \tilde{p}_0^2} \right] L \Bigg{ \}},
\end{aligned}  
\label{sieq:phi_BH}
\end{equation}

where
\begin{equation}
\begin{aligned}
\mu &= m c^2, \\
r_e &= \frac{e^2}{4 \pi \epsilon_0 m_e c^2},  \\
k &= h \nu = \tilde{E}_0 - \tilde{E} \Rightarrow \tilde{E}_0 = \tilde{E} + k, \\
\epsilon &= \ln \left( \frac{\tilde{E} + \tilde{p}}{\tilde{E} - \tilde{p}} \right) = 2 \ln \frac{\tilde{E} + \tilde{p}}{\mu}, \\
\epsilon_0 &= \ln \left( \frac{\tilde{E}_0 + \tilde{p}_0}{\tilde{E}_0 - \tilde{p}_0} \right) = 2 \ln \frac{\tilde{E}_0 + \tilde{p}_0}{\mu},\\
L &= \ln \left( \frac{\tilde{p}_0^2 + \tilde{p}_0 \tilde{p} - \tilde{E}_0 k}{\tilde{p}_0^2 - \tilde{p}_0 \tilde{p} - \tilde{E}_0 k} \right) \\
   &= 2 \ln \left( \frac{\tilde{E}_0 \tilde{E} + \tilde{p}_0 \tilde{p} - \mu^2}{\mu k} \right),  
\end{aligned}
\end{equation}
and $\tilde{p}=c p$, where $p$ is the electron momentum, and $\tilde{E} = E + \mu$, where $E$ is the kinetic energy of the electron. From the energy-momentum relation, we have $\tilde{E}^2 = \tilde{p}^2 + \mu^2 \Rightarrow \tilde{p} = \sqrt{\tilde{E}^2 - \mu^2}$, and $\tilde{p}_0 = \sqrt{\tilde{E}_0^2 - \mu^2}$. Additionally, using the above relation $\tilde{E}_0 = \tilde{E} + k$ we get  $\tilde{p} = \sqrt{(\tilde{E}_0 - k)^2 - \mu^2}$. Therefore, integrating over all incoming electron energies will give the $\Phi$ value at a specific frequency.

In our isotropic plasma, the electron distribution follows the relativistic Maxwellian distribution, with a momentum distribution of
\begin{equation}
  f (p) = \frac{e^{- \gamma (p) / \varepsilon}}{(m_ec)^3\varepsilon K_2 (1 / \varepsilon)},
\end{equation}
where $\gamma(p) = [1 + p^2/(m_ec)^2]^{1 / 2}$ and $\varepsilon = \kappa T/\mu$. Also, the electron incident velocity is $v = p_0 c^2/E_0$. Combining the above expressions for $E$, and $p_0$, the $\Phi$ in \cref{sieq:phi_BH} becomes a function of $k$ and $E_0$ only, i.e. $\Phi \equiv \Phi (E_0, k)$.

The total power spectrum at a specific frequency can be expressed as the integral over all $e$-$i$ collisions, i.e., integrating over all incident energies $E_0$. Thus,
\begin{equation}
  P (k) = \int_k^{\infty} \Phi (E_0, k) v f (p_0) p_0^2 \dd p_0.
\end{equation}
Using $c^2 p_0^2 = E_0^2 - \mu^2$, we can write $2 p_0 c^2 \dd p_0 = 2 E_0 \dd E_0 \Rightarrow \dd p_0 = v^{-1} \dd E_0$. Therefore,
\begin{equation}
  P (k) = \int_k^{\infty} \Phi (E_0, k) f (p_0) p_0^2 d E_0.
\end{equation}
To find the total power, we can then integrate $P (k)$ over all frequencies to
get
\begin{equation}
  P = \int_0^{\infty} P (k) \dd k.
\end{equation}

The \gls{bh} model gives the number spectrum of emitted photons in units of m$^3$/s/J, and thus 
\begin{equation}
  \frac{\dd I^2}{\dd \Omega \dd \omega} \Bigg|_{ei} = \dfrac{n_e n_i k \hbar}{4\pi} P(k),
\end{equation}
which is given in units of J/m$^3$/sr.

\subsection{Electron-electron bremsstrahlung}

According to the \gls{h} model, the number of photons emitted from an electron-electron gas in thermal equilibrium and uniform number density $n_e$ per unit time, per unit volume, and per unit energy interval is given by \cite{Haug1989}
\begin{equation}
    P(k,\tau) = \dfrac{1}{2} n_e^2 c \int p_1 f(\bm{p_1}) \int p_2 f(\bm{p_2}) \dfrac{\sqrt{\tilde{\mu}^2-1}}{\epsilon_1 \epsilon_2}\dfrac{\dd \sigma}{\dd k} \dd^3 p_1 \dd^3 p_2,
    \label{sieq:Haug_Pktau}
\end{equation}
where $\epsilon_1$ and $\epsilon_2$ are the total energies of the electrons and $\dd \sigma/\dd k$ is the cross section of two electrons leading to the generation of a photon of energy $k$. Also, $\bm{p_1}$ and $\bm{p_2}$ are the momenta of the two colliding electrons.

The integral in \cref{sieq:Haug_Pktau} is simplified to \cite{Haug1989}
\begin{equation}
\begin{aligned}
    P_{e e} (k, \tau) = &\frac{n_{\mathrm{e}}^2 c}{4 \tau [K_2 (1 / \tau)]^2}
    \int_1^{\infty} (\tilde{\mu} - 1)  \sqrt{\frac{\tilde{\mu} + 1}{2}} \\
    & \int_0^{k_{\mathrm{max}}^{\ast}}  \frac{d
    \sigma^{\ast} (k^{\ast}, \tilde{\mu})}{dk^{\ast}} \exp \left( - \frac{\sqrt{2 (\tilde{\mu}
    + 1)}}{\tau}  \frac{k^2 + k^{\ast 2}}{2 kk^{\ast}} \right) \frac{\dd k^{\ast}}{k^{\ast}}  \dd \tilde{\mu},
\end{aligned}
\end{equation}
where $k$ is the photon energy generated from the collision of the two electrons, and ${k }^{\mathord{*}}$ is the same energy measured from the center-of-mass system of the two colliding electrons. Also, $\tau = \kappa T / (m_e c^2)$, and 
\begin{equation}
    k_{\max}^{\ast} = \frac{\tilde{\mu} - 1}{\sqrt{2 (\tilde{\mu} + 1)}},
\end{equation}
where $d \sigma^{\ast} / d k^{\ast}$ is the e-e cross section. The second integral is split into
\begin{equation}
  \begin{aligned}
  &\int_0^{k_{\mathrm{max}}^{\ast}} \frac{dk^{\ast}}{k^{\ast}}  \frac{d
  \sigma^{\ast} (k^{\ast}, \tilde{\mu})}{dk^{\ast}} \exp \left( - \frac{\sqrt{2 (\tilde{\mu} +
  1)}}{\tau}  \frac{k^2 + k^{\ast 2}}{2 kk^{\ast}} \right) = \\
  & \int_0^{0.4 (\varepsilon - 1)} \frac{dk^{\ast}}{k^{\ast}}  \left[ \frac{d
   \sigma^{\ast}}{dk^{\ast}} \right]_2 \exp (- \ldots) + \int_{0.4
   (\varepsilon - 1)}^{k_{max}^{\ast}} \frac{dk^{\ast}}{k^{\ast}}  \left[
   \frac{d \sigma^{\ast}}{dk^{\ast}} \right]_1 \exp (- \ldots)
   \label{sieq:ee_brem_temp_1}
  \end{aligned}
\end{equation}

The two $d \sigma^{\ast} / d k^{\ast}$ terms are given using the following equations \cite{Haug1989}:
\begin{equation}
  \begin{aligned}
    \left[ \frac{d \sigma}{dk} \right]_2 & \approx \frac{\alpha r_0^2}{k} 
    \left\{ \ln \frac{4 \varepsilon p^2}{k}  \left[ \frac{32}{3} + \frac{8}{3
    \varepsilon^2 p^2} - \frac{(\varepsilon^2 + p^2)^3}{\varepsilon^4 p^4} +
    \frac{(\varepsilon^2 + p^2)^2}{\varepsilon^5 p^5} \ln (\varepsilon + p)
    \right] - \frac{16}{3} + \frac{4}{\varepsilon^2} - 2
    \frac{\varepsilon^2}{p^4} \right.\\
    & + \frac{2}{\varepsilon p}  \left( 1 + \frac{3}{\varepsilon^2} +
    \frac{\varepsilon^4}{p^4} \right) \ln (\varepsilon + p) + \frac{J
    (\varepsilon)}{\varepsilon p^2}  \left( \frac{1}{\varepsilon^2} - 4 -
    \frac{1}{2 \varepsilon^2 p^2} \right) \\
    &+ k \left[ \ln \frac{4 \varepsilon
    p^2}{k} \left\{ \frac{1}{\varepsilon} \right( - \frac{32}{3} + \frac{1}{6
    p^2} + \frac{1}{\varepsilon^2 p^4} - \frac{3 \varepsilon^4}{4 p^6} -
    \frac{3 p^2}{4 \varepsilon^4} \right)\\
    & \left. + \frac{1}{4 p} \ln (\varepsilon + p)  \left( \frac{1}{p^4} -
    \frac{1}{\varepsilon^4} + \frac{3}{p^6} + \frac{3}{\varepsilon^6} \right)
    \right\} - \frac{16}{3}  \frac{p^2}{\varepsilon^3} - \frac{2}{\varepsilon
    p^2} + \frac{2}{3 \varepsilon p^4} - \frac{1}{2 \varepsilon^5} -
    \frac{1}{\varepsilon p^6}\\
    & + \frac{1}{\varepsilon^2 p} \ln (\varepsilon + p)  \left( - 2 +
    \frac{3}{\varepsilon^2} - \frac{1}{\varepsilon^2 p^2} - \frac{1}{2
    \varepsilon^4} + \frac{1}{2 \varepsilon^2 p^4} + \frac{1}{p^6} \right) -
    \frac{J (\varepsilon)}{2 \varepsilon^2 p^4}  \left( 1 + \frac{3}{4
    \varepsilon^2} + \frac{3}{4 p^2} \right) \Bigg{]}\\
    & + k^2  \left[ \ln \frac{4 \varepsilon p^2}{k}  \left\{ \frac{49}{4 p^2}
    + \frac{1}{6 p^4} - \frac{29}{24 p^6} + \frac{7}{8 p^8} - \frac{17}{4
    \varepsilon^2} - \frac{37}{12 \varepsilon^4} + \frac{5}{8 \varepsilon^6} +
    \frac{1}{\varepsilon p} \ln (\varepsilon + p) \right. \right.\\
    & \left. \left( \frac{7}{4 \varepsilon^2} - \frac{1}{2 \varepsilon^4} +
    \frac{5}{8 \varepsilon^6} - \frac{39}{4 p^2} + \frac{1}{4 p^4} +
    \frac{5}{8 p^6} - \frac{7}{8 p^8} \right) \right\} \\
    &+ \frac{4}{3
    \varepsilon^2} + \frac{7}{2 \varepsilon^4} - \frac{9}{16 \varepsilon^6} -
    \frac{29}{48 \varepsilon^2 p^4} - \frac{5}{4 \varepsilon^4 p^4} -
    \frac{6}{p^6} - \frac{31}{16 \varepsilon^2 p^8}\\
    & + \frac{1}{\varepsilon p} \ln (\varepsilon + p)  \left( 2
    \frac{\varepsilon^4}{p^8} - \frac{\varepsilon^2}{16 p^8} + \frac{19}{8
    p^6} + \frac{4 \varepsilon^2}{p^4} + \frac{9}{4 \varepsilon^4 p^4} -
    \frac{2}{\varepsilon^2} - \frac{9}{16 \varepsilon^6} \right)\\
    & + \frac{J (\varepsilon)}{\varepsilon p^2}  \left( 2 - \frac{33}{32 p^2}
    - \frac{9}{4 p^4} - \frac{23}{32 p^6} - \frac{15}{32 \varepsilon^2} +
    \frac{9}{32 \varepsilon^4} \right) \Bigg{]} \Bigg{ \}}
  \end{aligned}
\end{equation}
and 
\begin{equation}
  \begin{aligned}
    \left[ \frac{d \sigma}{d k} \right]_1 = & \frac{\alpha r_0^2}{\varepsilon
    pk}  \left\{ \sqrt{\frac{p^2 - \varepsilon k}{\varepsilon^2 - \varepsilon
    k}}  \left[ \frac{1}{p} \ln (\varepsilon + p) \left\{ 2 \varepsilon +
    \frac{1}{\varepsilon} + (\varepsilon - k) \left( 1 + \frac{4 - 2
    \varepsilon k}{p^2} + \frac{1}{p^4} + \frac{k}{\varepsilon p^2 + k}
    \right) \right. \right. \right. & \\
    & \left. + \frac{p^2}{p^2 - \varepsilon k}  \left( 4 k - \varepsilon +
    \frac{2 k}{p^2 - \varepsilon k} \right) \right\} - (\varepsilon - k) 
    \left( \frac{16}{3} \varepsilon + 2 \frac{2 \varepsilon - k}{p^2} +
    \frac{1}{\varepsilon p^4} \right) - 4 k^2 - \frac{k}{\varepsilon} & \\
    & - \frac{2 p^2}{p^2 - \varepsilon k} \huge{]} + \frac{k}{2 p}  \sqrt{p^2
    - \varepsilon k} \ln \frac{2 p \sqrt{\varepsilon^2 - \varepsilon k} + k}{2
    p \sqrt{\varepsilon^2 - \varepsilon k} - k}  \left[
    \frac{\varepsilon}{\varepsilon p^2 + k} - \frac{\varepsilon^2 + p^2}{p^2 -
    \varepsilon k} - \frac{2 \varepsilon k}{(p^2 - \varepsilon k)^2} \right] &
    \\
    & - \frac{\varepsilon^2 k^2}{2 p \sqrt{(p^2 - \varepsilon k) 
    (\varepsilon^2 - \varepsilon k)}} \ln \frac{2 \varepsilon p + (\varepsilon
    - p) k}{2 \varepsilon p - (\varepsilon + p) k} \left( 2 \frac{\varepsilon
    - k}{p^2 - \varepsilon k} + \frac{\varepsilon^2 + 1}{\varepsilon p^2 + k}
    \right) & \\
    & + \left( 18 - \frac{2 k}{\varepsilon} + \frac{1}{\varepsilon^2} \right)
    L_1 + \frac{L_3}{\varepsilon p}  \left[ \frac{32}{3} \varepsilon^3
    (\varepsilon - k) + 8 \varepsilon^2 k^2 + k^2 - 14 \varepsilon^2 -
    \frac{14}{3} p^2 + \frac{19}{3} \varepsilon k \right. & \\
    & - \frac{k}{\varepsilon} - \frac{k}{p^2}  (\varepsilon - k) - \frac{k}{2
    \varepsilon p^4} - \frac{\varepsilon^2 + p^2}{p^2 - \varepsilon k} +
    \frac{1}{p} \ln (\varepsilon + p)  \{ 16 \varepsilon p^2 - 10 p^2 k - 4
    \varepsilon - k  & \\
    & \left. + \frac{11}{2 \varepsilon} + \frac{4 k - \varepsilon}{p^2} +
    \frac{k}{p^4} + \frac{3 (\varepsilon^2 + p^2) - 4 \varepsilon^3 k}{2 (p^2
    - \varepsilon k)} k + \frac{p^2}{2}  \frac{\varepsilon + k}{(p^2 -
    \varepsilon k)^2} - 2 \frac{(\varepsilon^2 + p^2)^3}{\varepsilon^2 k}
    \right\} & \\
    & + \frac{1}{p} \ln \frac{2 p^2 - (\varepsilon - p) k}{2 p^2 -
    (\varepsilon + p) k}  \left\{ 2 \varepsilon^3 - 4 \varepsilon + k +
    \frac{3}{4 \varepsilon} \right. & \\
    & \left. + \frac{1}{p^2 - \varepsilon k}  \left( \varepsilon^3 k^2 -
    \frac{3}{4} k (\varepsilon^2 + p^2) - \frac{p^2}{4} \left.
    \frac{\varepsilon + k}{p^2 - \varepsilon k} \right) \right\} \right] & \\
    & + \frac{2 L_1 L_3}{p \sqrt{(\varepsilon^2 - \varepsilon k)  (p^2 -
    \varepsilon k)}}  \left[ 6 p^2 k - 16 \varepsilon p^2 - \varepsilon k^2 -
    2 \varepsilon - \frac{3}{\varepsilon} + \frac{(\varepsilon^2 +
    p^2)^3}{\varepsilon^2 k} \right] & \\
    & + \frac{k^2}{\sqrt{\varepsilon^2 - \varepsilon k}}  \int_{- 1}^{+ 1} d
    (\mathrm{cos} \theta)  \left[ \frac{2 L}{\sqrt{R_1}}  \left\{ \frac{2 -
    \varepsilon^2}{\kappa_2} + \frac{2}{\kappa_1^2} + \frac{\varepsilon
    p^2}{\kappa_1 \kappa_2}  (4 \varepsilon - k) \right. \right. & \\
    & + \frac{4 p^2 - 3 \varepsilon k + 1 / \kappa_1}{\kappa_1  (2 p^2 -
    \kappa_2)} + \frac{\varepsilon}{\kappa_2 R_1}  [\varepsilon \kappa_1 -
    (\varepsilon - k) \kappa_2]  \left( \frac{2 p^2}{\kappa_1} +
    \frac{\varepsilon^2 + 1}{\varepsilon^2 - \varepsilon k} \right)  \Bigg\}  & \\
    & + \frac{L_2}{W_2}  \left\{ 2 \frac{\varepsilon^2 + p^2}{\kappa_2} -
    \frac{3}{2} + \frac{\varepsilon^2 + p^2 - 2 \varepsilon k}{\varepsilon k} 
    \left( \varepsilon^2 - \varepsilon k + \frac{\kappa_1}{4} \right) +
    \frac{p^2}{2 p^2 - \kappa_2} \right. & \\
    & +\frac{1}{\kappa_1  (2 p^2 - \kappa_2)}  \Bigg[(\varepsilon^2 + p^2 - 2
    \varepsilon k) \{ (\varepsilon^2 + p^2 - \varepsilon k) \kappa_2 -
    (\varepsilon^2 + p^2) \}  \left. - \kappa_2 - \frac{\kappa_2^2}{2} \right]
    & \\
    & + \frac{1}{4 \varepsilon k (2 p^2 - \kappa_2)}  \Bigg[2 (\varepsilon^2 + p^2
    - 2 \varepsilon k) \{ \varepsilon k + 1 - 4 p^2 (p^2 - \varepsilon k) \} 
    & \\
    & + \kappa_1^2 - 3 \kappa_1 + (\varepsilon^2 + p^2) \kappa_2  \Bigg]  \Bigg\} -
    \frac{L_4}{W_4} \left\{ 1 - \frac{\varepsilon k}{2 p^2} - \frac{1}{4 p^4}
    \right. + \frac{1}{\kappa_1 \kappa_2}  (1 - \kappa_1 - \kappa_1^2) & \\
    &  \left. + \frac{\varepsilon^2 + p^2 - 2 \varepsilon k}{\kappa_1
    \kappa_2}  \left[ 4 p^2  (p^2 - \varepsilon k) + \varepsilon^2  \left( 2
    k^2 - 2 + \frac{\varepsilon k}{p^2} \right) \right] \right\} \Bigg]  \Bigg\},
  \end{aligned}
\end{equation}
where

\begin{equation}
\begin{array}{ll}
     & \kappa_1 = k (\varepsilon - p \cos \theta),\\
     & \kappa_2 = k (\varepsilon + p \cos \theta),\\
     & L_1 = \ln \left( \sqrt{\varepsilon^2 - \varepsilon k} + \sqrt{p^2 -
     \varepsilon k} \right),\\
     & L_2 = \ln \left\{ 1 + 2 \frac{\varepsilon - k}{k \kappa_1} 
     \left[ (p^2 - \varepsilon k) \kappa_2 + \sqrt{p^2 - \varepsilon k} W_2
     \right] \right\},\\
     & L_3 = \ln \frac{\left( \varepsilon \sqrt{p^2 - \varepsilon k} + p
     \sqrt{\varepsilon^2 - \varepsilon k} \right)^2}{\varepsilon k},\\
     & L_4 = \ln \left\{ 1 + 2 \frac{\varepsilon^2 - \varepsilon k}{\kappa_1
     \kappa_2}  \left[ 4 p^2 (p^2 - \varepsilon k) + \sqrt{p^2 - \varepsilon
     k} W_4 \right] \right\},\\
     & L = \ln \left\{ \frac{\sqrt{\varepsilon^2 - \varepsilon k}}{\kappa_1} 
     \left[ 2 p^2 - \kappa_2 + \sqrt{(p^2 - \varepsilon k) R_1} \right]
     \right\},\\
     & R_1 = 4 (p^2 - \varepsilon k) + 4 \kappa_1 +
     \frac{\kappa_1^2}{\varepsilon^2 - \varepsilon k},\\
     & W_2 = \sqrt{\frac{\varepsilon \kappa_2}{\varepsilon - k}  [2 k^2 +
     (p^2 + k^2 - 2 \varepsilon k) \kappa_2]},\\
     & W_4 = 2 p \sqrt{4 p^2  (p^2 - \varepsilon k) + \frac{\kappa_1
     \kappa_2}{\varepsilon^2 - \varepsilon k}},\\
     & \alpha \approx 1 / 137,\\
     & r_0 = e^2 / (m c^2),
   \end{array}
\end{equation}
and
\begin{equation}
\begin{aligned}
J (\epsilon) &= \int_{- 1}^{+ 1} \frac{\ln (\epsilon -
   px)}{\epsilon + px} dx \\
   &= \frac{1}{p}  \sum_{n = 1}^{\infty} \frac{1}{n}
   \left\{ \ln (\epsilon + p)  \left[ \left( \frac{\epsilon + p}{2
   \epsilon} \right)^n + \left( \frac{\epsilon - p}{2 \epsilon}
   \right)^n \right] - \frac{1}{n}  \left[ \left( \frac{\epsilon + p}{2
   \epsilon} \right)^n - \left( \frac{\epsilon - p}{2 \epsilon}
   \right)^n \right] \right\}.  
\end{aligned}
\end{equation}

In addition,  $\tilde{\mu} = \epsilon^2 + p^2$ and $p^2 = \epsilon^2 - 1$; thus, $\varepsilon = \sqrt{(\tilde{\mu} + 1)/2}$. Therefore, the middle integration point becomes $0.4 (\varepsilon - 1) = 0.4 \left( \sqrt{(\tilde{\mu} + 1)/2} - 1 \right)$
Ref. \cite{Haug1989} also notes that applying the following correction to the cross section equations improves accuracy
\begin{equation}
  F_{e e} = \frac{a_2}{a_1} \frac{e^{2 \pi a_1} - 1}{e^{2 \pi a_2} - 1},
\end{equation}
where
\begin{equation}
   a_1 = \frac{\varepsilon^2 + p^2}{2 \varepsilon p} \alpha 
\end{equation}
and
\begin{equation}
    a_2 = \frac{\varepsilon^2 + p^2 - 2.0 \varepsilon k}{2 \sqrt{(\varepsilon^2
   - \varepsilon k) (p^2 - \varepsilon k)}} \alpha.
\end{equation}

To increase the numerical stability of the code, we compute the lower $k^{\mathord{*}}$ limit. The exponential term in \cref{sieq:ee_brem_temp_1} goes to zero when the argument goes to $-x$, where $x \gg 1$. We set $C = \sqrt{2 (\tilde{\mu} + 1)}/\tau$ and thus
\begin{equation}
   - C \frac{k^2 + k^{\ast 2}}{2 kk^{\ast}} \geqslant - x ,
\end{equation}
\begin{equation}
   k^2 - \frac{x}{C} 2 kk^{\ast} + k^{\ast 2} \le 0 .
   \label{sieq:temp_11}
\end{equation}
Next we set $G = 2 x/C = 2 x \tau/\sqrt{2 (\tilde{\mu} + 1)}$, and so \cref{sieq:temp_11} becomes $k^{\mathord{*} 2} - G k k^{\mathord{*}} + k^2 \le 0$. This is a quadratic equation, and thus we can solve the equality to get the two roots
\begin{equation}
k^{\mathord{*}}_{1, 2} = \frac{G k \pm k \sqrt{G^2 - 4}}{2} ,
\end{equation}
which are used to restrict the limits of the integral.

To convert the above number of photons emitted per m$^3$ per s per dimensionless photon energy $k / (m_e c^2)$ to J/m$^3$/sr we do 
\begin{equation}
  \frac{\dd I^2}{\dd \Omega \dd \omega}\Bigg|_{ee} = \frac{k \hbar}{4\pi m_e c^2} P_{e e}. 
\end{equation}

\subsection{Total bremsstrahlung}

The total relativistic bremsstrahlung power is
\begin{equation}
  \frac{\dd I^2}{\dd \Omega \dd \omega} = \frac{\dd I^2}{\dd \Omega \dd \omega}\Bigg|_{e i} + \frac{\dd I^2}{\dd \Omega \dd \omega}\Bigg|_{e e}.
\end{equation}

\begin{figure}
    \centering
    \includegraphics[width=0.5\linewidth]{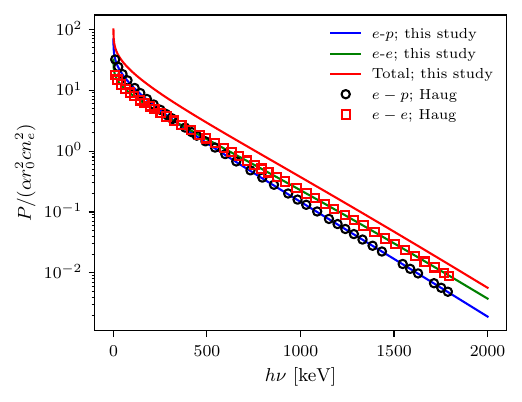}%
    \includegraphics[width=0.5\linewidth]{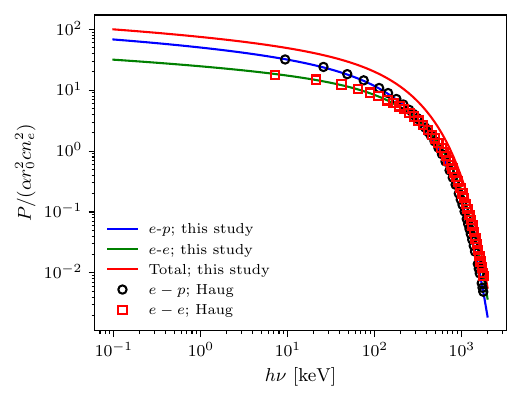}
    \caption{Bremsstrahlung code validation. The comparison results are obtained from Ref. \cite{Haug1989}.}
    \label{sifig:brem_spectra_haug_comparison}
\end{figure}

\Cref{sifig:brem_spectra_haug_comparison} shows the validation of the current code compared to the data presented in the \gls{h} model paper \cite{Haug1989}.

\begin{figure}[!htb]
    \centering
    \includegraphics[width=1\linewidth]{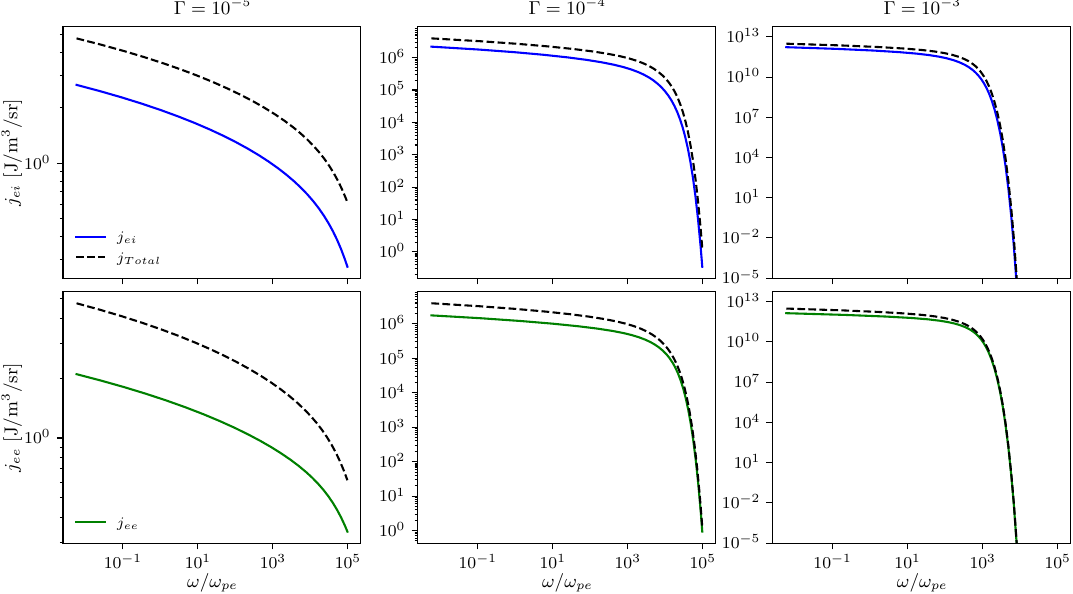}
    \caption{Theoretical relativistic bremsstrahlung for electron-ion and electron-electron collisions for $\varepsilon\approx1$.}
    \label{subfig:analytical_brem_eps1}
\end{figure}

\begin{figure}[!htb]
    \centering
    \includegraphics[width=1\linewidth]{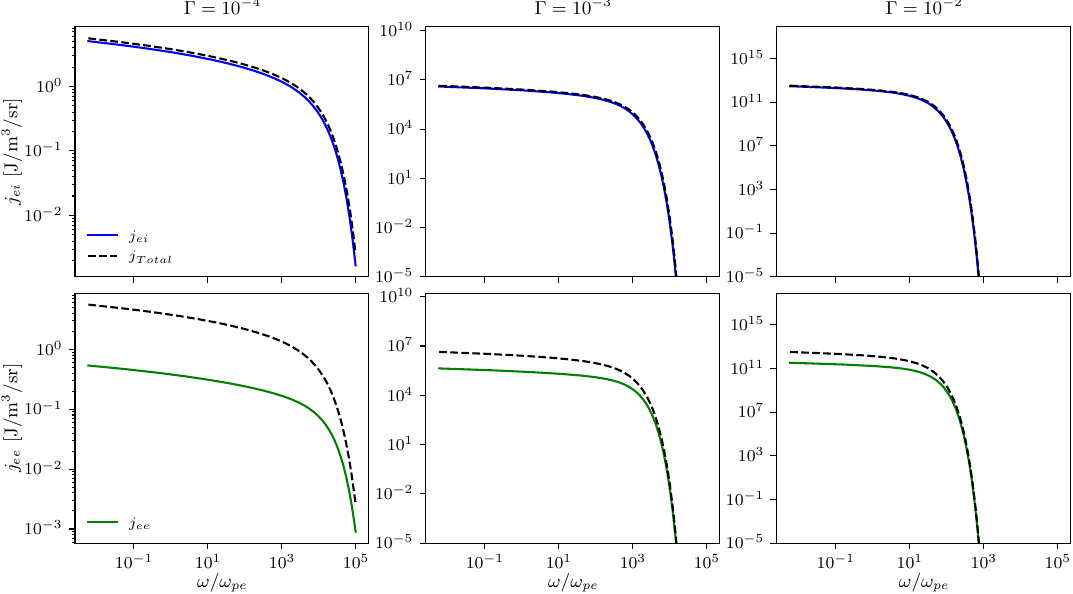}
    \caption{Theoretical relativistic bremsstrahlung for electron-ion and electron-electron collisions for $\varepsilon\approx0.1$.}
    \label{subfig:analytical_brem_eps0p1}
\end{figure}

\Cref{subfig:analytical_brem_eps1,subfig:analytical_brem_eps0p1} shows the theoretical bremsstrahlung from the \gls{bh} and \gls{h} models for different relativistic parameters. The code to compute the \gls{bh} and \gls{h} emission spectra is provided in the supplementary material of the main paper.

\section{Theoretical synchrotron radiation}
\label{sisec:th_sync}

Based on the \gls{s} model \cite{Schwinger1949}, the spectrum of an electron moving with constant speed in a circular path (periodic motion) consists of harmonics of the rotational angular frequency $\omega_0 = v / R$. The power radiated into the $\xi$th harmonic, integrated over all angles, is \cite{Schwinger1949}
\begin{equation}
  P_{\xi} = \frac{\xi \omega_0 e^2}{4 \pi \epsilon_0 R} \left[ 2 \beta^2 J_{2
  \xi}' (2 \xi \beta) - (1 - \beta^2) \int_0^{2 \xi \beta} J_{2 \xi} (x) d x
  \right],
\end{equation}
where
\begin{equation}
\begin{aligned}
  \omega = \xi \omega_0, \\
  \omega_0 = \frac{\beta c}{R}, \\
  R = \frac{\beta E}{e c B},\\
  E = \gamma m_e c^2.
\end{aligned}
\end{equation}
Combining the above gives $\omega_0 = e B/\gamma m_e$. The specific $\gamma$ that is required to excite a specific harmonic $\xi$ is
\begin{equation}
  \gamma = \xi \frac{e B}{\omega m_e}.
\end{equation}

The $P_{\xi}$ above is in units of J/s, and thus 
\begin{equation}
  \frac{d^2 I}{d \Omega d \omega} = \frac{\tau}{4 \pi V} P_{\xi},
\end{equation}
gives units of W/m$^3$/Hz/sr.

The above analysis was done for 1 electron. Therefore, for our isotropic plasma, we need to integrate
over the MJ distribution
\begin{equation}
f(\beta) = \dfrac{\gamma^5 \beta^2}{\varepsilon \textrm{K}_2(1/\varepsilon)}e^{-\gamma/\varepsilon}.
\label{sieq:mj_dist_vel}
\end{equation}
Thus,
\begin{equation}
  \frac{d^2 I}{d \Omega d \omega} = \frac{1}{4 \pi} \int_0^{2 \pi}
  \int_0^{\pi} \int_0^1 \frac{d^2 I (\beta)}{d \Omega d \omega} f (\beta) d
  \beta \sin \theta d \theta d \phi,
\end{equation}
or equivalently
\begin{equation}
  \frac{d^2 I}{d \Omega d \omega} = \int_0^1 \frac{d^2 I (\beta)}{d \Omega d
  \omega} f (\beta) d \beta.
\end{equation}
To compute $d \beta$ we start from
\begin{equation}
  \frac{1}{\sqrt{1 - \beta^2}} = \xi \frac{\omega_c}{\omega},
\end{equation}
where $\omega_c = e B / m_e$, and thus
\begin{equation}
  | d \beta | = \frac{\omega}{\beta \xi^2 \omega_c^2} | d \omega |.
\end{equation}

To find the emission spectra in units of J/m$^3$/sr, we multiply by the number of electrons $N_e$, i.e., 
\begin{equation}
  \frac{d^2 I}{d \Omega d \omega} = N_e \left[ \frac{d^2 I}{d \Omega d  \omega} \right]_e.
\end{equation}

\FloatBarrier
\section{Additional simulation results}
\label{sisec:extra_sim_results}

\begin{figure}[htb]
    \centering
    \includegraphics[width=1\linewidth]{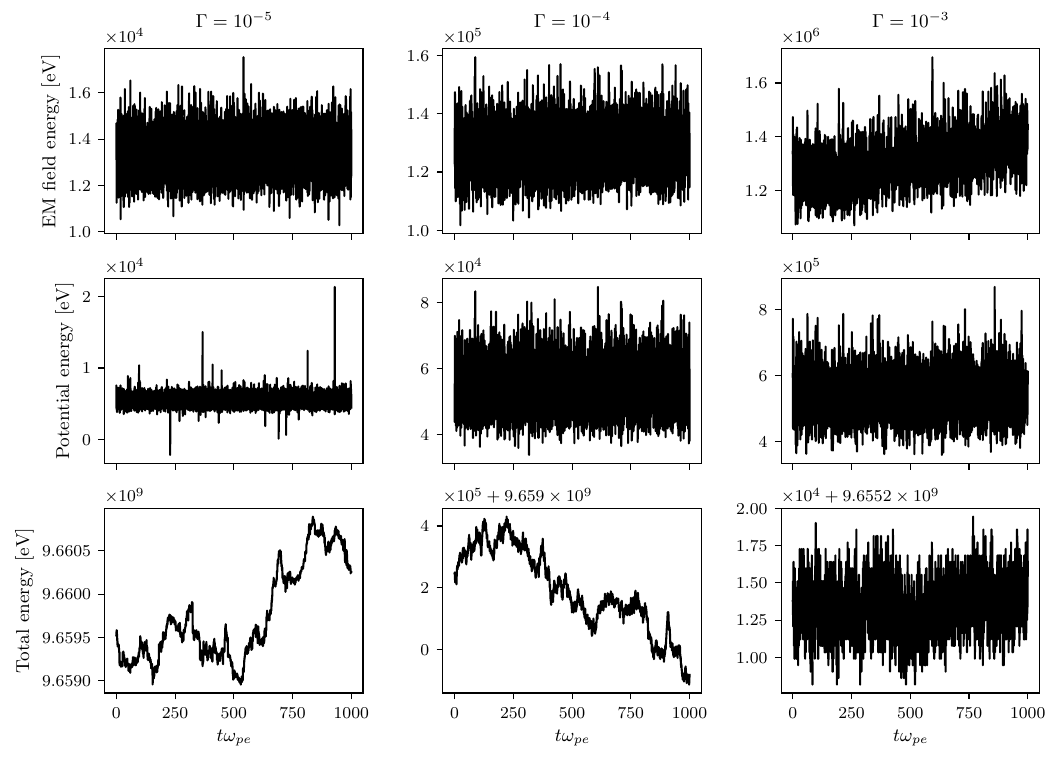}
    \caption{Energies from \gls{md} simulations at different $\Gamma$ conditions and at $\varepsilon\approx1$.}
    \label{sifig:lammps_energies}
\end{figure}

\begin{figure}[htb]
    \centering
    \includegraphics[width=0.5\linewidth]{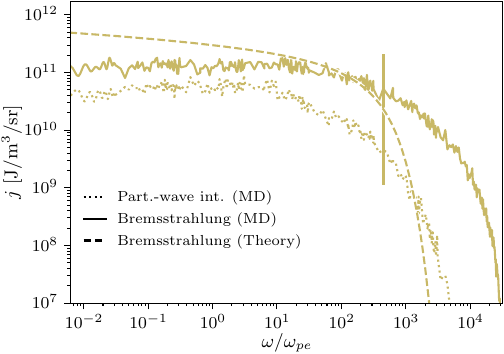}%
    \includegraphics[width=0.5\linewidth]{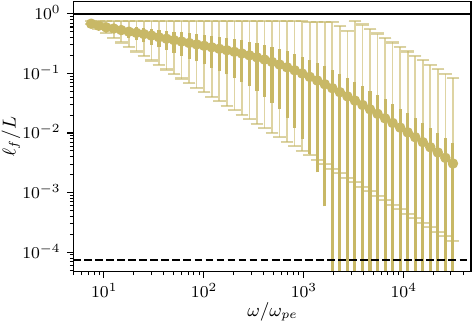}

    \includegraphics[width=0.5\linewidth]{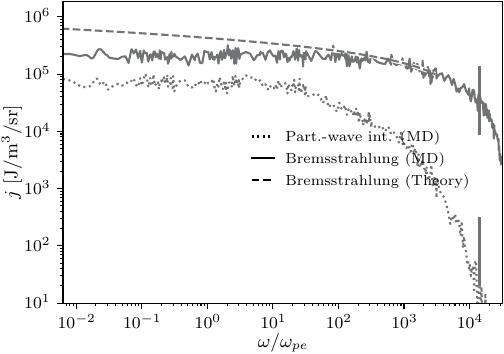}%
    \includegraphics[width=0.5\linewidth]{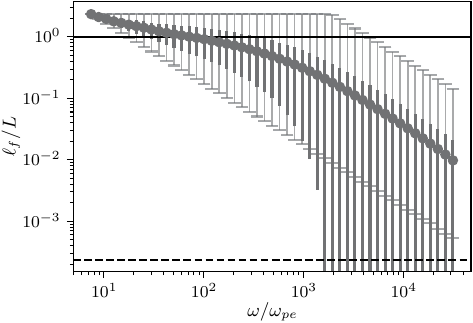}

    \includegraphics[width=0.5\linewidth]{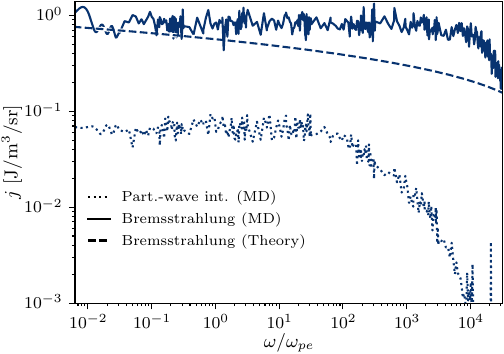}%
    \includegraphics[width=0.5\linewidth]{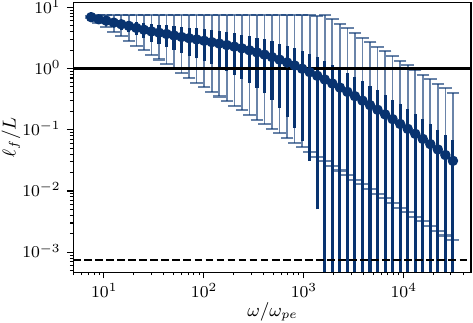}

    \caption{Emission spectra (left) and photon formation lengths (right) computed from simulations with case numbers  (cf. Table~I) 8 (top), 5 (middle), and 1 (bottom). In the right-side graphs, the circle markers denote the average over all electrons, the thick line error bars without caps show the standard deviation, and the thin line error bars with caps correspond to the minimum and maximum from all events measured. Additionally, the horizontal dashed black line corresponds to $\ell_{f,min}/L$, where $\ell_{f,min}=c \Delta t$.}
    \label{sifig:l_f_sim}
\end{figure}

\FloatBarrier

\subsection{Radiation Reaction}
\label{sisec:rad_reac}

Below, we investigate whether radiation reaction (RR) is significant within our simulations. The ratio of the RR damping force to the Lorentz force is given by the parameter \cite{Nielsen2021,Gonoskov2022}
\begin{equation}
\eta=\alpha_f\gamma\psi=\dfrac{\alpha_f\gamma^2}{E_0}\sqrt{(\bm{\mathcal{E}}+\bm{v}\times \bm{B})^2-\left(\dfrac{\bm{v} \cdot \bm{\mathcal{E}}}{c}\right)^2},
\end{equation}
where $\alpha_f$ is the fine-structure constant, $E_0=m_e^2/(e\hbar)\simeq1.32\times10^{18}$ V/m the critical field, and $\psi$ the strong-field parameter. The RR force dominates the dynamics for $\eta\approx10$ to $100$ \cite{Nielsen2021} and quantum effects become important for $\psi \ge 1$ \cite{Nielsen2021}. Considering that electrons experience both electric and magnetic fields in our simulations, the upper bound of the field term is 
\begin{equation}
\sqrt{(\bm{\mathcal{E}}+\bm{v}\times \bm{B})^2-(\bm{v} \cdot \bm{\mathcal{E}})^2}\le |\bm{\mathcal{E}}+\bm{v} \times \bm{B}|\le \mathcal{E}+cB \le \mathcal{E}_{\rm max}+c B_{\rm max},
\end{equation}
where the worst-case scenario is investigated, i.e., using the strongest electric and magnetic fields found in the domain and using an electron's velocity $\rightarrow c$. Thus,
\begin{equation}
\eta_{\max}=\dfrac{\alpha_f\gamma^2}{E_0} (\mathcal{E}_{\max}+c B_{\max}).
\end{equation}
For the $T=500$ keV simulations in Fig. 1 of the manuscript, $\langle\gamma^2\rangle=13.6$ from the the Maxwell--J\"uttner distribution and thus $\eta_{\max}\approx 6\times10^{-9}$, $6\times10^{-7}$ and $5\times10^{-5}$ for $\Gamma=10^{-5}$, $10^{-4}$ and $10^{-3}$, respectively. Thus, $\eta\ll 10$ for all conditions, concluding that RR is negligible in our simulations.

\FloatBarrier
\bibliography{references}